\documentclass[%
graphicx,
aip,
jmp,
amsmath,
amssymb,
preprint,%
]{revtex4-1}

\usepackage{dcolumn}
\usepackage{bm}

\begin{document}

\title[{\footnotesize A.M. Escobar-Ruiz, J.C. L\'opez Vieyra and P.
Winternitz}]{Fourth order
superintegrable systems separating in Polar Coordinates. I.  Exotic Potentials}

\author{Adrian M. Escobar-Ruiz}
\email{escobarr@crm.umontreal.ca}
\affiliation{
Centre de recherches math\'ematiques,
and D\'epartement de math\'ematiques \\
et de statistique, Universit\'e de Montreal,  C.P. 6128,
succ. Centre-ville,\\ Montr\'eal (QC) H3C 3J7, Canada}
\author{J. C. L\'opez Vieyra}%
\email{vieyra@nucleares.unam.mx}
\affiliation{ Instituto de Ciencias Nucleares, Universidad Nacional
Aut\'onoma de M\'exico, Apartado Postal 70-543, 04510 M\'exico, D.F., Mexico
}%

\author{P. Winternitz}
\email{wintern@crm.umontreal.ca}
\affiliation{
Centre de recherches math\'ematiques,
and D\'epartement de math\'ematiques \\
et de statistique, Universit\'e de Montreal,  C.P. 6128,
succ. Centre-ville,\\ Montr\'eal (QC) H3C 3J7, Canada}

\date{\today}

\begin{abstract}
We present all real quantum mechanical potentials in a two-dimensional Euclidean
space that have the following properties: 1. They allow separation of variables of the
Schr\"odinger equation  in polar coordinates, 2. They allow an independent fourth
order integral of
motion, 3. It turns out that their angular dependent part $S(\theta)$ does not satisfy any linear equation. In this case $S(\theta)$ satisfies a nonlinear ODE that has the Painlev\'e
property and its solutions can be expressed in terms of the
Painlev\'e transcendent $P_6$. We also study the corresponding classical analogs of these potentials. The polynomial algebra of the integrals of motion is constructed in the classical case.
\end{abstract}


\maketitle

\section{INTRODUCTION}

This article is part of a series devoted to a study of classical and quantum superintegrable systems. Roughly speaking, a Hamiltonian with $n$ degrees of freedom is integrable if it allows $n$ independent well defined integrals of motion in involution.
It is minimally superintegrable if it allows $n+1$ such integrals, maximally superintegrable if it allows $2n-1$ integrals (only subsets of $n$ integrals among them can be in involution).

The best known superintegrable systems are the harmonic oscillator with its $su(n+1)$ algebra of integrals, and the Kepler-Coulomb system with its $o(n+1)$ algebra (when restricted to fixed bound state energy values).

A recent review article gives more precise definitions, a general setting and
motivation for studying superintegrable
systems~\cite{MillerPostWinternitz:2013}. It follows from Bertrand's
theorem~\cite{Bertrand}  that $\omega r^2$ and $\alpha/r$ are the only
maximally
superintegrable spherically symmetrical potentials in Euclidean real space $E_n$.
Systematic searches for superintegrable classical and quantum  systems in $E_2$
and $E_3$ established a connection between second order superintegrability and
multiseparability in the Schr\"{o}dinger or Hamilton-Jacobi
equation~\cite{Capel:2015,Fris:1965,Fris:1966,Makarov:1967,KalninsMW:1976,
Carinena:2017}.

An extensive literature exists on second order superintegrability in spaces of
$2$, $3$ and $n$ dimensions, Riemannian and pseudo-Riemannian, real or complex,
\cite{Kalnins:2002, KalninsI:2005, KalninsII:2005, KalninsIII:2005, KalninsIV:2006,
KalninsV:2006}.

A systematic study of higher order integrability is more recent. Pioneering work
is due to Drach~\cite{Drach:1935,Drach:1935b}.  For more recent work
see~\cite{Bermudez:2016,Chanu:2011,Chanu:2012,Hietarinta:1998,Tsiganov:2000,GravelWinternitz:2002,Marquette:2009,
Marquette:2010, Ranada:2013,Celeghini:2013,Fernandez:2016,Hakobyan:2016,GungorKNN:2014,TTW:2009,TTW:2010,QuesneTTW:2010,
PostVinetZhedanov:2011,MarquetteSajediW:2017}.

The Painlev\'e transcendents were first introduced in a purely mathematical study by Painlev\'e\cite{painleve1902} and Gambier\cite{gambier1910} and were popularized in books e.g. by Ince\cite{Ince} and Davis\cite{Davis}.  They are characterized by the fact that they are solutions of second order nonlinear ODEs that are single valued about any movable singularity of the ODE (movable means that the position of the singularity depends on the initial conditions). We shall call this “the Painlev\'e property”. Painlev\'e and Gambier also classified all ODEs with the Painlev\'e property of the form  $y'' = R(x,y,y’)$  with $R$ rational in $y$ and $y'$ and analytical in $x$ into $50$ equivalence classes under the action of the group preserving the Painlev\'e property. Of these $50$ six give rise to the famous irreducible Painlev\'e transcendents. The others can either be reduced to one of these six, or integrated in terms of already known functions like elliptic functions, or solutions of linear equations.
 Linear ODEs have the Painlev\'e property by default: all the singularities of their solutions are fixed, i.e., they can only occur where the coefficients of the ODEs are themselves singular. In this sense we can say that the nonlinear ODEs with the Painlev\'e property are the closest ones to linear ODEs.
Nonlinear equations with the Painlev\'e property became important in applications after the discovery of the inverse scattering theory by Kruskal et al.\cite{Kruskal} and more generally of soliton theory (for reviews see e.g. \cite{Ablowitz,Conte,Miura} and references therein). A Painlev\'e test was proposed \cite{Ablowitz2,Ablowitz3,Ablowitz4}, a simple algorithmic test the passing of which is a necessary condition for an ODE to have the Painlev\'e property.  A Painlev\'e conjecture was formulated\cite{Ablowitz2}, namely that a necessary condition for a PDE to be integrable by inverse scattering techniques is that all of the ODE’s obtained as reductions of the PDE should have the Painlev\'e property.
A systematic search for analytical solutions of many of the PDEs  of  hydrodynamics, plasma physics, and nonlinear optics lead to various Painlev\'e transcendents. Painlev\'e transcendents to our knowledge appeared for the first time in quantum mechanics in articles by Fushchych and Nikitin\cite{Nikitin} and by Doebner and Zhdanov\cite{Doebner}.
A systematic search for superintegrable sxystems   in $E_2$ with one integral of motion of order $N\geq3$ and two others of order $N \leq2$ was started in \cite{Gravel:2004} and \cite{GravelWinternitz:2002} (for $N=3$). Exotic potentials, by definition not satisfying any linear ODE, were obtained. It turned out that they could always be expressed in terms of the Painlev\'e transcendents $P_1, P_2, P_4$ or elliptic functions. The lower order integrals were chosen to be of “Cartesian type” that is they forced the potential to allow separation of variables in Cartesian coordinates. A similar study for $N=3$ was conducted for second order integrals of polar type \cite{TremblayW:2010}. Exotic potentials appeared again and this time they were expressed in terms of $P_6$. For $N=4$ the situation is similar\cite{MarquetteSajediW:2017}, namely, exotic potentials appear in the Cartesian case, expressed in terms of $P_1 ,..., P_5$. For $N=4$, the polar case is the present article, and as we shall see below exotic potentials exist. Unlike the case $N=3$, they are expressed in terms of the completely general $P_6$ transcendent.  Specific results have also been obtained for $N=5$ in the Cartesian case\cite{Abouamal}. New features appear here, namely potentials expressed in terms of solutions of higher order ODEs with the Painlev\'e property.
 We conjecture that for all $N\geq3$ exotic potentials will exist and be solutions of ODEs with the Painlev\'e property.

The present article is a contribution to a
series~\cite{GravelWinternitz:2002,Gravel:2004,TremblayW:2010,
PostWinternitz:2015,
MarchesielloPostSnobl:2015,PopperPW:2012,MarquetteSajediW:2017} devoted to
superintegrable systems in $E_2$
with one integral of order $n\ge 3$ and one of order $n\leq 2$. In particular,
it is a generalization of a paper\cite{TremblayW:2010} devoted to the case of a third order
integral $Y$.

In this article we restrict ourselves to the space $E_2$.  The Hamiltonian has
the form
\begin{equation}
\label{H0}
H\ =\ \frac{1}{2}(p_x^2+p_y^2) + V(x,y) \ ,
\end{equation}
in classical mechanics $p_x$ and $p_y$ are the momenta conjugate to the
Cartesian coordinates $x$ and $y$. In quantum mechanics they are the
corresponding operators
$p_{x} = -i \hbar \frac{\partial}{\partial x}$, $p_{y} = -i \hbar
\frac{\partial}{\partial y}$. In {polar coordinates} $(x,y)\equiv
(r\cos\theta,\,r\sin\theta)$, the classical Hamiltonian reads

\begin{equation}
\label{H}
H\ =\ \frac{1}{2} \left(p_r^2+ \frac{p_\theta^2}{r^2}\right) +V(r,\theta)  \ ,
 \qquad V(r,\theta) = R(r) + \frac{1}{r^2} S(\theta)\ ,
\end{equation}
here $p_r$ and $p_\theta$ are the associated canonical momenta. The corresponding quantum operator takes the form
\begin{equation}
\label{Hpolar}
H\ =\ -\frac{\hbar^2}{2}\,\bigg(\partial^2_r + \frac{1}{r}\partial_r + \frac{1}{r^2}\partial_\theta^2\bigg)  + V(r,\theta) \ .
\end{equation}
In this article we concentrate on quantum superintegrability and on
"exotic" potentials, namely those that do not satisfy any linear differential
equations. In all equations we keep the Planck constant $\hbar$ explicitly.
Classical exotic potentials will be obtained in the limit $\hbar\rightarrow 0$.
We emphasize that this limit is singular: highest order terms in the equation
which defines the potential in (\ref{H0}) vanish, so the classical and quantum
cases can differ greatly.

In addition to the Hamiltonian $H$, we have two more conserved quantities which are
 \begin{align}
\label{X}
 X&\ =\ p_\theta^2+2\,S(\theta)\ ,\\[10pt]
\label{Y}
 Y&\ =\  \! \!  \sum_{i+j+k=4} \! \!  A_{ijk}\,\{ L_z^i,\,p_x^j\,p_y^k \}\
+ \ \{ g_1(x,y),\,p_x^2  \} \ + \  \{ g_2(x,y),\,p_x\,p_y  \}
 \\ & \hspace{50pt}
\ + \  \{ g_3(x,y),\,p_y^2  \}
\ + \  g_4(x,y)\ ,\nonumber
 \end{align}
here $p_\theta=x\,p_y-y\,p_x=-i \hbar \frac{\partial}{\partial \theta}$. The bracket $\{\cdot ,\, \cdot   \}$ denotes an anticommutator, the set $\{A_{ijk}\}$ are real constants and $R(r),\, S(\theta),\, g_{1,2,3,4}(x,y)$ are real functions such that
\begin{equation}
[H,\,Y] \ = \ [H,\,X] \ = \ 0   \ .
\end{equation}
The operator $Y$ in (\ref{Y}) is given in Cartesian coordinates for brevity. Putting
\[
p_x \ = \ -i\,\hbar\,(\cos \theta \,\partial_r - \frac{\sin \theta}{r}\,\partial_\theta) \ , \qquad p_y \ = \ -i\,\hbar\,(\sin \theta \,\partial_r + \frac{\cos \theta}{r}\,\partial_\theta) \ ,
\]
we obtain the corresponding expression in polar coordinates. It's leading terms are given explicitly below in (\ref{Ylead}) and used throughout this article.

We have $[Y,X]=C \neq 0$ where $C$ is in general a $5$th order linear operator.
In general, we thus obtain a finitely generated polynomial algebra of integrals
of motion
\cite{Daskaloyannis:2006,Daskaloyannis:2007,KalninsI:2005,KalninsII:2005,Kalnins:2000b,Kalnins:2008,Marquette:2010,
MikiPVZ:2013}.
We are looking for fourth-order superintegrable systems, so at least one of
$A_{ijk}$ is different from zero. The operator $Y$
is the most general polynomial expression for a fourth-order Hermitian operator of the required form.
The commutator $[H,\,Y]$  contains derivatives of order up to three.

Before calculating the commutator $[H,Y]$ we note that three ''trivial'' fourth
order integrals exist, namely $X^2$, $H^2$ and $\{X, H\}$. Each of these is a
scalar (invariant) under $O(2)$ rotations. By linear combinations of the form $Y
+ u_1\, X^2 + u_2\, H^2 + u_3\, X  H$, where the $u_i$ are constants, we
can eliminate 3 parameters among the $\{A_{ijk}\}$ and consequently three terms
in $Y$. Now, we introduce a more convenient set of parameters defined by the
relations
\begin{equation}
\begin{aligned}
A_{0 0 4} &= \frac{1}{2} \, (D_{1} - B_{1})\,,
\quad \ \ A_{0 4 0} = \frac{1}{2} ( B_{1} + D_{1})\,,
\qquad \ \, A_{4 0 0} = 0\,,
 \\
A_{0 2 2} &= -3 \,D_{1} \,,
\hspace{1.4cm} \ \ A_{0 1 3} = B_{2} - 2\, D_{2}\,,
\hspace{1.3cm} A_{0 3 1} = B_{2} + 2\, D_{2}\,,
 \\
A_{1 0 3} &=     A_{4} -\,C_1,
\qquad \quad \ A_{3 0 1} =  A_{2}\,,
\hspace{2.5cm} A_{2 2 0} =  B_{3}\,,
 \\
A_{1 2 1} &=  3\,C_1+A_4 \,,
\hspace{1.05cm} A_{1 1 2} =  A_{3} -  3\,C_{2}\,,
\qquad \ \quad A_{3 1 0} = A_{1}\,,
 \\
A_{2 1 1} &= 2\,B_{4}\,,
\hspace{1.9cm} A_{1 3 0} = C_2+A_3\ ,
\hspace{1.5cm} A_{2 0 2} = -B_{3} \ .
\end{aligned}
\end{equation}
With the above parameters the fourth-order integral (\ref{Y}) takes
the following form:

\begin{equation}
\begin{aligned}
\label{YA}
Y &\ =\  A_1\,\{L_z^3,\ p_x \} \ + \ A_2\,\{L_z^3,\ p_y \} \ + \ A_3\,\{L_z,\
p_x\,(p_x^2+p_y^2) \} \ + \ A_4\,\{L_z,\ p_y\,(p_x^2+p_y^2)    \} \\ &
+ \ B_1\,(p_x^4-p_y^4) \ + \ 2\,B_2\,p_x\,p_y\,(p_x^2+p_y^2) \ + \ B_3\,
\{L_z^2,\  p_x^2-p_y^2  \} \ + \ 2\,B_4\,\{L_z^2,\ p_x\,p_y \}
\\ &
+ \ C_1\,\{L_z,\ 3\,p_x^2\,p_y-p_y^3 \} + C_2\,\{L_z,\ p_x^3 -3\,p_y^2\,p_x \} \
+ \ D_1\,(p_x^4+p_y^4-6\,p_x^2\,p_y^2) \ + \
\\ &
4\,D_2\,p_x\,p_y\,(p_x^2-p_y^2)
+ \text{lower order terms}\ .
\end{aligned}
\end{equation}
Under rotations around the z-axis, each of the six pairs of parameters
\[
(A_1,A_2),\ (A_3,A_4),\ (B_1,B_2),\ (B_3,B_4),\ (C_1,C_2),\ (D_1,D_2)\ ,
\]
in (\ref{YA}) forms a doublet (all $O(2)$ singlets have been removed).  Under
rotations through the angle $\theta$ the doublets $A_i,\,B_i,\,C_i$ and $D_i$
rotate through $\theta,\,2\theta,\,3\theta$ and $4\theta$, respectively. In
particular, the doublets $(A_1,\,A_2)$ and $(B_3,\,B_{4})$ will play a central
role in the main equations of the present paper. Explicitly, in polar
coordinates, the leading terms of the integral $Y$ are
\begin{equation}
\begin{aligned}
\label{Ylead}
Y &\ = \ \hbar^4\,\bigg( ( B_1 \cos 2 \theta + B_2 \sin 2 \theta + D_1 \cos 4 \theta   + D_2
\sin 4 \theta)\,\partial^4_r
+  \frac{1}{ r^4}  \bigg[   D_2 \sin 4 \theta
\\ & + D_1 \cos 4 \theta  -2 r \left(A_{1} r^2+A_4\right) \sin \theta
  - \left(B_2+2\,B_{4} r^2\right) \sin 2 \theta +2 r \left(A_2 r^2+A_4\right)
\cos \theta
\\ &
 - \left(B_1+2\,B_3 r^2\right) \cos 2 \theta  -2\,r\, (
  \,C_1\,  \cos 3 \theta-\,C_2\,  \sin 3 \theta)    \bigg]\,  \partial^4_\theta \\ &
- \frac{2}{r^2} \bigg[
3(
  D_{1} \cos 4\theta
+ D_{2} \sin 4\theta
)
- r^2 (
  B_{3}   \cos 2\theta
+ B_{4}  \sin 2\theta )
\\ &
+ r   (
A_{3} \sin \theta - A_4 \cos  \theta
-3(\,C_1 \, \cos 3\theta
- \,C_2 \sin 3\theta)\,  )
\bigg] \partial^2_r\,\partial^2_\theta
\\ &  - \frac{2}{ r} \bigg[   B_1 \sin 2 \theta- B_2 \cos 2 \theta +2 (D_1 \sin
4 \theta  - D_2 \cos 4 \theta)
 -r\,(\,C_1\, \sin 3 \theta + \,C_2\, \cos 3 \theta
\\ &  + A_3 \cos \theta + A_4 \sin \theta      )
\bigg]\,\partial^3_r\,\partial_\theta
-\frac{2}{r^3}  \bigg[    B_1 \sin 2 \theta- B_2 \cos 2 \theta
\\ &   -2 \,(D_1 \sin 4 \theta + D_2 \cos 4 \theta)
  - r (A_3 \cos \theta+ \,A_4 \sin \theta   -3 ( \,C_1\,
\sin 3 \theta + C_2\, \cos 3 \theta ) \ )
\\ &     + 2\,r^2 \left( B_3 \sin 2 \theta - B_{4} \cos 2 \theta \right) -\, r^3
\left( A_{1} \cos \theta +A_2 \sin \theta \right)
  \bigg]\, \partial_r\,\partial^3_\theta \ \bigg) + \ldots +\text{lower order terms}  \ .
\end{aligned}
\end{equation}

We introduce the functions
\[
G_1(r,\,\theta) \ = \ g_1\,\cos^2\theta +
g_2\,\sin^2\theta+g_3\,\cos\theta\sin\theta \ ,
\]
\[
G_2(r,\,\theta) \ = \ \frac{
g_1\,\sin^2\theta+ g_2\,\cos^2\theta-g_3\,\cos\theta\sin\theta}{r^2} \ ,
\]
\[
G_3(r,\,\theta) \ = \ -\frac{
g_1\,\sin2\theta-g_2\,\sin2\theta-g_3\,\cos2\theta}{r} \ ,
\]
\begin{equation}
\label{Gfunctionsg}
G_4(r,\,\theta) \ = \ g_4 \ .
\end{equation}
Then, the quadratic and zero order terms in the integral $Y$ can now be written in polar
coordinates as
\begin{equation}
\begin{split}
& \{ g_1(x,y),\,p_x^2  \} \ + \  \{ g_2(x,y),\,p_x\,p_y  \}
\ + \  \{ g_3(x,y),\,p_y^2  \}
\ + \  g_4(x,y)\
\\&
=   \ -\hbar^2\,(\{ G_1(r,\theta),\,\partial_r^2  \} \ + \  \{ G_3(r,\theta),\,\partial_r\,\partial_\theta  \}
\ + \  \{ G_2(r,\theta),\,\partial_\theta^2  \})
\ + \  G_4(r,\theta)\,.
\end{split}
\end{equation}

The structure of this article is as follows. In Section \ref{CommutatorHY} we derive the determining equations that govern the existence and form of the
fourth-order integral $Y$. In \hbox{Section \ref{linearcompatibility}} we present a linear
compatibility condition that must be satisfied by the potential $V(r,\theta)$ in order for a fourth order integral $Y$ to exist. In general this is a fourth order PDE.  In Section \ref{SUPERINTEGRABILITY} we turn to the question of superintegrability.
The existence of the second order integral $X$ guarantees that the potential  $V(r,\theta)$ has the form given in (\ref{H}). We rewrite the determining equations and the compatibility condition (\ref{CC}) in polar coordinates. The compatibility condition (\ref{CC}) then reduces to a coupled system of ODEs for $R(r)$ and $S(\theta)$. We decouple the equation and solve for $R(r)$. The possible functions $R(r)$ are $br^2, a/r$ and $0$ respectively.  From Section \ref{Potentials} on we restrict to exotic potentials which by definition do not satisfy any linear equation. The function $R(r)$ is already determined and is not exotic. The function $S(\theta)$ satisfies a linear equation which must be satisfied identically. This requires that all coefficients in $Y$ vanish except $A_1,A_2$ $B_3$ and $B_4$. In Section \ref{Potentials} we consider the case $R(r)=0$, {\it i.e.} a nonconfining potential (with no bound states).  Section \ref{CONFINING POTENTIALS}  is devoted to confining potentials $R(r)=b r^2$ and $R(r)= \frac{a}{r}$.  In all cases the function $S(\theta)$ is expressed in terms of the Painlev\'e transcendent $P_6 (\gamma_1, \gamma_2, \gamma_3,\gamma_4;z)$ where  $\gamma_1, \gamma_2, \gamma_3,\gamma_4$ are arbitrary constants. Section \ref{CLASSICAL POTENTIALS} is devoted to classical potentials obtained in the (singular) limit $\hbar \to 0$. The fourth order compatibility condition reduces to a second order non-linear ODE which, interestingly,  does not have the Painlev\'e property.  The polynomial algebra generated by the integrals of motion is presented in Section \ref{POLYNOMIAL ALGEBRA}. The main results are summed up as theorems in the final Section \ref{Conclusions}.

\section{DETERMINING EQUATIONS FOR A FOURTH ORDER INTEGRAL}
\label{DETERMINING EQUATIONS}

\subsection{Commutator $[H,\,Y]$}
\label{CommutatorHY}
The commutator between the Hamiltonian $H$ (\ref{Hpolar}) and the fourth order integral $Y$, written in polar coordinates, is a third
order differential operator given by
{
\begin{align}
[H,Y]   &=
{\cal A}_{rrr}\,\frac{\partial^3  }{ \partial r^3}\  + \
{\cal A}_{rr\theta}\,\frac{\partial^3  }{\partial r^2\partial\theta}\ + \
{\cal A}_{r\theta\theta}\,\frac{\partial^3  }{\partial r\partial \theta^2}\ + \
{\cal A}_{\theta\theta\theta}\,\frac{\partial^3  }{ \partial \theta^3}
\nonumber \\&
+{\cal B}_{rr}\,\frac{\partial^2  }{\partial r^2}\ + \
{\cal B}_{r\theta}\,\frac{\partial^2  }{\partial r\partial\theta}\ + \
{\cal B}_{\theta\theta}\,\frac{\partial^2  }{\partial \theta^2}
\nonumber \\&
 +{\cal C}_{r}\,\frac{\partial  }{\partial r}\ + \
 {\cal C}_{\theta}\,\frac{\partial  }{\partial\theta} \ + \
  {\cal C}_{0}  = 0\  ,
\label{commutator}
\end{align}
}
where the coefficients ${\cal A}_{rrr}, {\cal A}_{rr\theta},...$, are real
functions of $r$ and $\theta$. Terms multiplying derivatives of order five and
four vanish identically (they are already accounted for in the form of $Y$ in
(\ref{Y})). In order for $Y$   to be an integral of
motion all ten coefficients must vanish simultaneously. The odd order terms in
(\ref{commutator}) provide  us with useful information.
The even order terms
${\cal B}_{rr},\,{\cal B}_{\theta\theta},\ {\cal B}_{r \theta}$ and ${\cal
C}_{0}$ provide differential  consequences of the odd order terms
and will not be listed
below. This difference between  even and odd order terms in the commutator $[H,Y]$ is a general feature of the theory\cite{PostWinternitz:2015} (for any order $N$ of $Y$).

Vanishing of the coefficients of the third order terms ${\cal A}_{rrr}$, ${\cal
A}_{\theta\theta\theta}$, ${\cal A}_{rr\theta}$,
${\cal A}_{r\theta\theta}$ in (\ref{commutator}) yields, respectively,  the
following relations:

\begin{align}
 G_{1}^{(1,0)} & \, = \,  {F}_1(\theta)\, {V}^{(1,0)}  +  {F}_2(r,\theta)\,
{V}^{(0,1)}\,,
\label{Eq30}  \\
 \frac{1}{r^2}
 \Big(
G_{2}^{(0,1)} + \frac{1}{r} G_{3}
  \Big) & \, = \,  {F}_3(r,\theta)\, {V}^{(1,0)} +  {F}_4(r,\theta)\,
{V}^{(0,1)}\,,
\label{Eq03} \\
\frac{1}{r^2}G_{1}^{(0,1)} +  G_{3}^{(1,0)}
& \, =  \,  3\,{F}_2(r,\theta)\, {V}^{(1,0)} +  {F}_5(r,\theta)\, {V}^{(0,1)}
\,,
\label{Eq21} \\
  \frac{2}{r^3} G_1
+ G_{2}^{(1,0)}
+ \frac{1}{r^2} G_{3}^{(0,1)}
& \,  =  \, {F}_5(r,\theta)\, {V}^{(1,0)} +  3\,{F}_3(r,\theta)\, {V}^{(0,1)}
\, ,
 \label{Eq12}
\end{align}

From  the two equations ${\cal C}_{r} = 0$ and ${\cal C}_{\theta}=0$, we obtain

\begin{equation} 
\begin{aligned}
&  2 \,G_1 \,V^{(1,0)} + G_3 \,V^{(0,1)}-\frac{1}{2}\,G_4{}^{(1,0)}
\\ &  + \hbar^2\bigg[
 \frac{G_1{}^{(0,2)}}{2 r^3}
-\frac{G_1{}^{(1,0)}}{r^2}
+\frac{G_1{}^{(1,2)}}{2 r^2}
+\frac{2 G_1{}^{(2,0)}}{r}
+G_1{}^{(3,0)}
-\frac{3}{2} G_2{}^{(1,0)}      \\&
+\frac{1}{2} G_2{}^{(1,2)}
-\frac{1}{2} r G_2{}^{(2,0)}
+\frac{G_3{}^{(0,3)}}{4 r^2}
-\frac{3\, G_3{}^{(0,1)}}{4 r^2}
+\frac{5 G_3{}^{(1,1)}}{4 r}
+\frac{3}{4} G_3{}^{(2,1)}
\bigg]
\\[10pt]&\hspace{30pt}
= \ \hbar ^2\bigg[F_{1}\,V^{(3,0)}  +  F_{3}\,V^{(0,3)}   +  F_{5}\,V^{(1,2)} +
F_{6}\,V^{(2,1)}
\\ &   +   F_{10}\,V^{(2,0)} +   2\,F_7\,V^{(1,1)}  +  F_8\,V^{(0,2)}  +
F_{11}\,V^{(1,0)}  +  F_{12}\,V^{(0,1)}   \bigg]
\ ,
\label{Eq10}
\end{aligned}
\end{equation}

\medskip

\begin{equation} 
\begin{aligned}
&
  2 \,G_2 \,V^{(0,1)}
    + G_3 \,V^{(1,0)}
-\frac{G_4{}^{(0,1)}}{2\, r^2}
+\hbar^2\bigg[
 \frac{G_1{}^{(1,1)}}{r^3}
+\frac{G_1{}^{(2,1)}}{2 r^2}
-\frac{G_2{}^{(0,1)}}{r^2}
+\frac{G_2{}^{(0,3)}}{r^2}        \\&
+\frac{1}{2} G_2{}^{(2,1)}
+\frac{G_3}{4 r^3}
+\frac{5 G_3{}^{(0,2)}}{4 r^3}
-\frac{G_3{}^{(1,0)}}{4 r^2}
+\frac{3 G_3{}^{(1,2)}}{4 r^2}
+\frac{G_3{}^{(2,0)}}{2 r}
+\frac{1}{4} G_3{}^{(3,0)}
\bigg]
\\[10pt]&\hspace{30pt}
   = \  \hbar ^2\bigg[  F_{2}\,V^{(3,0)}   +  3\,F_{3}\,V^{(1,2)}  +
F_{4}\,V^{(0,3)}  + F_{5}\,V^{(2,1)}
\\ &    +   F_{7}\,V^{(2,0)}  +  F_9\,V^{(0,2)}  +   2\,F_8\,V^{(1,1)} +
F_{12}\,V^{(1,0)}  +  F_{13}\,V^{(0,1)}  \bigg]
\ ,
\label{Eq01}
\end{aligned}
\end{equation}
respectively, where we define $V^{(i,j)}\equiv
\partial^i_r\,\partial^j_\theta\,V(r,\,\theta)$.

\clearpage

The Planck constant $\hbar$ is present in the lowest order coefficients ${\cal
C}_{r}$ and ${\cal C}_{\theta}$ only (see (\ref{Eq10}) and (\ref{Eq01})).

The functions $F_1 \ldots F_{13}$ are completely determined by the constants
$A,B,C,D$ figuring in the leading part of the integral $Y$. They are given in Appendix A.

\subsection{The linear compatibility condition}
\label{linearcompatibility}

The system  (\ref{Eq30})-(\ref{Eq12}) viewed as a system of 4 PDE for $G_1$, $G_2$
and $G_3$ is overdetermined and for general potential $V(r,\,\theta)$ has no
solutions. The first step towards finding solutions of this system is to
establish a necessary linear compatibility condition involving $V(r,\,\theta)$
alone. Such an equation will exist as a consequence of the equality of all mixed
derivatives of analytical functions.
To obtain the compatibility condition we denote the l.h.s. of the equations
(\ref{Eq30})-(\ref{Eq12}) as $E_1,
\ldots E_4$, respectively, and take partial derivatives of these terms (up to
third order). The following linear combination of the derivatives vanishes
identically
\begin{equation}
\begin{split}
0\ = \  &
{r}^{6}    {E_2} ^{(3,0)}
+12\,  {r}^{5}  {E_2}^{(2,0)}
+36\, {r}^{4}  {E_2}^{(1,0)}
- {r}^{4}   {E_4}^{(2,1)}
- {r}^{3}  {E_3}^{(2,0)}
\\&
-6\, {r}^{3}   {E_4}^{(1,1)}
+24\, {r}^{3} {E_2}
-3\, {r}^{2}  {E_3}^{(1,0)}
-6\,{r}^{2}   {E_4}^{(0,1)}
+ {r}^{2}   {E_3}^{(1,2)}
 \\&
  +2\, r  {E_3}^{(0,2)}
  +3\, r   {E_1}^{(1,1)}
  - {E_1}^{(0,1)}
  - {E_1}^{(0,3)}
\end{split}
\end{equation}
hence the same combination of the r.h.s. of (\ref{Eq30})-(\ref{Eq12}) must
vanish too and we obtain the compatibility condition:

\begin{equation}
\begin{aligned}
& 0 \ =  \
       r^6 F_3                        \, V^{(4,0)}
 +     r^4 \left(F_4 r^2 - F_5\right) \, V^{(3,1)}
 -  3\,r^2 \left(F_3 r^2 - F_2\right) \, V^{(2,2)} \\&
 +  (F_5 r^2 - F_1)                   \, V^{(1,3)}
 -   F_2                             \, V^{(0,4)}
+ r^3 \left( 3 r^3 F_3{}^{(1,0)}  + 12 r^2 F_3   - r F_5{}^{(0,1)} -3\,F_2
\right)\, V^{(3,0)}
\\&
 +  r \bigg(
    3  r^5 \, F_4{}^{(1,0)}
   +12 r^4 \, F_4
   -3  r^3 \, F_3{}^{(0,1)}
   -2  r^3 \, F_5{}^{(1,0)}
   -7  r^2 \, F_5
   +6  r   \, F_2{}^{(0,1)}
   +3 F_1\bigg)\, V^{(2,1)}
\\&
+ \bigg(
3 r^2    \, F_2{}^{(1,0)}
- 6  r^4    \, F_3{}^{(1,0)}
- 18 r^3    \, F_3
+  2 r^2    \, F_5{}^{(0,1)}
+ 9  r      \, F_2
- 3         \, F_1'
\bigg)\, V^{(1,2)}
\\&
+ \bigg( r^2 \, F_5{}^{(1,0)} + 2 r \, F_5   - 3 F_2{}^{(0,1)} \bigg)\,
V^{(0,3)}
+ r \bigg(
   3 r^5 \, F_3{}^{(2,0)}
+ 24 r^4 \, F_3{}^{(1,0)}
+ 36 r^3 \, F_3            \\&
- 2  r^3 \, F_5{}^{(1,1)}
- 6  r^2 \, F_5{}^{(0,1)}
- 6  r^2 \, F_2{}^{(1,0)}
+ 3  r   \, F_2{}^{(0,2)}
- 9  r   \, F_2
+ 3      \, F_1'
\bigg) \,V^{(2,0)}
\nonumber
\end{aligned}
\end{equation}

\begin{equation}
\begin{aligned}
&
+ \bigg(
   3 r^6 \, F_4{}^{(2,0)}
+ 24 r^5 \, F_4{}^{(1,0)}
- 6  r^4 \, F_3{}^{(1,1)}
+ 36 r^4 \, F_4
-    r^4 \, F_5{}^{(2,0)}
-18  r^3 \, F_3{}^{(0,1)}  \\&
-8   r^3 \, F_5{}^{(1,0)}
+6   r^2 \, F_2{}^{(1,1)}
+    r^2 \, F_5{}^{(0,2)}
- 9  r^2 \, F_5
+15  r   \, F_2{}^{(0,1)}
-  F_1
-3 F_1''
\bigg) \, V^{(1,1)}         \\&
-  \bigg(
+ 3 r^4 \, F_3{}^{(2,0)}
+18 r^3 \, F_3{}^{(1,0)}
+18 r^2 \, F_3
- 2 r^2 \, F_5{}^{(1,1)}
- 3 r   \, F_2{}^{(1,0)}
-4  r   \, F_5{}^{(0,1)}    \\&
+ F_2
+3 F_2{}^{(0,2)}
\bigg)\, V^{(0,2)}
+ \bigg(
     r^6 \, F_3{}^{(3,0)}
+ 12 r^5 \, F_3{}^{(2,0)}
+ 36 r^4 \, F_3{}^{(1,0)}
-    r^4 \, F_5{}^{(2,1)}    \\&
- 3  r^3 \, F_2{}^{(2,0)}
+ 24 r^3 \, F_3
- 6  r^3 \, F_5{}^{(1,1)}
- 9  r^2 \, F_2{}^{(1,0)}
+ 3  r^2 \, F_2{}^{(1,2)}
- 6  r^2 \, F_5{}^{(0,1)}    \\&
+ 6  r   \, F_2{}^{(0,2)}
-F_1' - F_1{}^{(3)}
\bigg) \, V^{(1,0)}
 + \bigg(
    r^6 \, F_4{}^{(3,0)}
+12 r^5 \, F_4{}^{(2,0)}
-3  r^4 \, F_3{}^{(2,1)}    \\&
+36 r^4 \, F_4{}^{(1,0)}
+24 r^3 \, F_4
-18 r^3 \, F_3{}^{(1,1)}
-   r^3 \, F_5{}^{(2,0)}
-18 r^2 \, F_3{}^{(0,1)}
-3  r^2 \, F_5{}^{(1,0)}    \\&
+   r^2 \, F_5{}^{(1,2)}
+2  r   \, F_5{}^{(0,2)}
+3  r   \, F_2{}^{(1,1)}
-F_2{}^{(0,1)}
-F_2{}^{(0,3)}
\bigg)  \, V^{(0,1)} \ .
\label{CC}
\end{aligned}
\end{equation}
Relation (\ref{CC}) is a fourth order linear PDE for the potential and is a
necessary (but not sufficient) condition for the existence of the fourth order
integral $Y$ of the form (\ref{Y}).
This relation does not contain the Planck constant $\hbar$ and is thus the same
in classical and in quantum mechanics.

\section{SUPERINTEGRABILITY: SEPARATION IN POLAR COORDINATES}
\label{SUPERINTEGRABILITY}

\subsection{The determining equations}

Vanishing of the commutator $[H,\,X]=0$ implies that the potential has the
separable form of $V(r,\,\theta)$ in (\ref{H}) and thus allows separation of
variables in polar coordinates in the Schr\"{o}dinger equation (and in the
Hamilton-Jacobi equation).

In this case the determining equations (\ref{Eq30})-(\ref{Eq01}), coming from
the condition $[H,\,Y]=0$, take the form
%
\begin{align}
 G_{1}^{(1,0)} & \, = \,     F_1\, R'  -\frac{2 \, F_1}{r^3}\,S   +  \frac{F_2
}{r^2}S'
\  ,
\label{Eq30p} \\
 \frac{1}{r^2}
 \Big(
G_{2}^{(0,1)}  + \frac{1}{r} G_{3}
  \Big) & \, = \,    F_3\, R'  -\frac{2 \, F_3}{r^3}\,S   + \frac{F_4 }{r^2}S'
\ ,
\label{Eq03p} \\
 \frac{1}{r^2}G_{1}^{(0,1)} +  G_{3}^{(1,0)}
& \, =  \,    3\,F_2\, R'  -\frac{6 \, F_2}{r^3}\,S      +\frac{F_5}{r^2}S'   \
,
\label{Eq21p} \\
  \frac{2}{r^3} G_1
+  G_{2}^{(1,0)}
+ \frac{1}{r^2} G_{3}^{(0,1)}
& \,  =  \,   F_5\, R'   -\frac{2 \, F_5}{r^3}\,S    +\frac{3\,F_3 }{r^2}S' \ ,
 \label{Eq12p}
\end{align}

\small{

\begin{equation}    
\begin{aligned}
\label{Eq10p}
&
      G_3 \,\frac{1}{r^2}\,S'
+ 2 \,G_1\, ( R'- \frac{2}{r^3}\,S) -\frac{1}{2} G_4{}^{(1,0)}
\\ &
+ \hbar^2\bigg[\frac{G_1{}^{(0,2)}}{2 r^3}-\frac{3 G_3{}^{(0,1)}}{4 r^2}
+\frac{G_3{}^{(0,3)}}{4 r^2}-\frac{G_1{}^{(1,0)}}{r^2}-\frac{3}{2} G_2{}^{(1,0)}
+\frac{5 G_3{}^{(1,1)}}{4 r}
\\ &
+\frac{G_1{}^{(1,2)}}{2 r^2}
+\frac{1}{2} G_2{}^{(1,2)}
+\frac{2 G_1{}^{(2,0)}}{r}
-\frac{1}{2} r G_2{}^{(2,0)}
+\frac{3}{4} G_3{}^{(2,1)}
+G_1{}^{(3,0)}\bigg] \\ &
\\&\hspace{20pt}
= \ \hbar^2\bigg[
 \left(\frac{6 F_6}{r^4}-\frac{24 F_1}{r^5}-\frac{2 F_{11}}{r^3}\right)\,S
 +\left(\frac{6 F_{10}}{r^4}-\frac{4 \,F_7}{r^3}+\frac{F_{12}}{r^2}\right) S'
\\ &  \hspace{30pt}
+\left(\frac{F_8}{r^2}-\frac{2 \,F_5}{r^3}\right)\, S''
+\frac{F_3}{r^2}\,S'''
+ F_{11}\, R'
+F_6\, R''
+ F_1 \,R'''
 \bigg]
\, ,
\end{aligned}
\end{equation}

\begin{equation}  
\begin{aligned}
\label{Eq01p}
& G_3 \,( R'- \frac{2}{r^3}\,S)   +2 \,G_2\,\frac{1}{r^2}\,S'
-\frac{G_4{}^{(0,1)}}{2 r^2}
\\ &
+\hbar^2\bigg[
 \frac{G_3}{4 r^3}
-\frac{G_2{}^{(0,1)}}{r^2}
+\frac{5 G_3{}^{(0,2)}}{4 r^3}
+\frac{G_2{}^{(0,3)}}{r^2}
-\frac{G_3{}^{(1,0)}}{4 r^2}
+\frac{G_1{}^{(1,1)}}{r^3}      \\ &
+\frac{3 G_3{}^{(1,2)}}{4 r^2}
+\frac{G_3{}^{(2,0)}}{2 r}
+\frac{G_1{}^{(2,1)}}{2 r^2}
+\frac{1}{2} G_2{}^{(2,1)}
+\frac{1}{4} G_3{}^{(3,0)}
\bigg]
\\&\hspace{30pt}
\\ &  = \  \hbar^2\bigg[
\left(\frac{6 \,F_7}{r^4}-\frac{24 \,F_2}{r^5}-\frac{2\, F_{12}}{r^3}\right)\,S
+\left(\frac{6\, F_5}{r^4}-\frac{4\, F_8}{r^3}+\frac{F_{13}}{r^2}\right)\, S'
\\ &  \hspace{30pt}
+\left(\frac{F_9}{r^2}-\frac{6 F_3}{r^3}\right) \,S''
+ \frac{F_4}{r^2}\, S'''
+ F_{12}\, R'
+  F_7 \,R''
+  F_2\,R'''
\bigg]
\ ,
\end{aligned}
\end{equation}

}

\subsection{The linear compatibility condition}

Substituting the separable form (\ref{H}) of the potential into the
compatibility condition (\ref{CC}), and integrating once over $r$, we obtain
{\small
\begin{equation}
\begin{aligned}
\Theta(\theta) & =
288\, \Big[
  B_{{1}}  \sin 2\,\theta
- B_{{2}}  \cos 2\,\theta
- 8 D_{{1}} \,\sin  4\,\theta
+ 8 D_{{2}} \,\cos  4\,\theta
\Big] \, S
 \\&
+120\, \Big[
   20 D_{{1}} \,\cos  4\,\theta
+ 20 D_{{2}} \,\sin  4\,\theta
- B_{{1}} \cos 2\,\theta
- B_{{2}} \sin 2\,\theta
\Big] \, S'
\\&
+60\, \Big[
  B_{{1}}  \sin  2\,\theta
-  B_{{2}} \cos  2\,\theta
+ 14 D_{{1}} \,\sin 4\,\theta
- 14 D_{{2}} \,\cos 4\,\theta
 \Big] \, S''
 \\&
 -30\, \Big[
  B_{{1}} \cos  2\,\theta
+ B_{{2}} \sin  2\,\theta
+ 4 D_{{1}}\,\cos  4\,\theta
+ 4 D_{{2}}\,\sin  4\,\theta
\Big] \, S^{(3)}
 \\&
-3\, \Big[
 B_{{1}} \sin  2\,\theta
- B_{{2}} \cos  2\,\theta
+2 D_{{1}} \,\sin 4\,\theta
-2 D_{{2}} \,\cos 4\,\theta
\Big]\,   S^{(4)}
 \\[5pt] & \hspace{-30pt} 
- r \Big(
96\, \Big[
     A_4 \,\sin \theta
+    A_3 \,\cos \theta
- 9\, C_1\, \sin  3\,\theta
- 9\, C_2\, \cos  3\,\theta
\Big] \, S
\\&
-120\, \Big[
  A_{{4}} \,\cos \theta
-  A_{{3}} \,\sin \theta
+ 9\,C_2\,\sin  3\,\theta
- 9\,C_1\,\cos  3\,\theta
\Big] \, S'
\\&
+480\, \Big[
\,C_1 \,\sin 3\,\theta
+\,C_2 \,\cos 3\,\theta
\Big] \,S''
\\ &
+30\, \Big[
A_{{3}}\,\sin  \theta
-A_{{4}}\,\cos  \theta
+ 3\,C_2 \,\sin 3\,\theta
- 3\,C_1 \,\cos 3\,\theta
\Big] \,  S^{(3)}
 \\ &
 - 6\,\Big[
 A_{{4}} \,\sin \theta
+ A_{{3}} \,\cos \theta
+ C_1 \,\sin  3\,\theta
+ C_2 \,\cos  3\,\theta
\Big] \,   S^{(4)}
\Big)
\\[5pt]&
+576\,{r}^{2} \Big[
   D_{{2}} \cos  4\,\theta
-  D_{{1}} \sin  4\,\theta  \Big] R
\\&
+6\,{r}^{3} \Big[
6\, r^3 \, \big(
 A_{{2}} \, \sin  \theta
+  A_{{1}}\, \cos  \theta
\big)
-4\,{r}^{2} \big(
   B_{{4}} \cos  2\,\theta
-   B_{{3}}  \sin  2\,\theta
\big)
 \\&
 -r\, \big(
 A_{{4}} \,\sin  \theta
+ A_{{3}} \,\cos  \theta
+ 9\, C_1 \,\sin  3\,\theta
+ 9\, C_2 \,\cos  3\,\theta
\big)
 \\&
-3  \,B_{{1}}\,\sin  2\,\theta
+3  \,B_{{2}}\,\cos  2\,\theta
+66 \, D_{{1}}\,\sin 4\,\theta
-66 \, D_{{2}}\,\cos 4\,\theta
 \Big] \, R'
  \\[5pt]&
  +6\,{r}^{4}
 \Big[
 6\,{r}^{3}\big(
 A_{{2}} \,\sin  \theta
+  A_{{1}} \,\cos  \theta
\big)
 - 4\,{r}^{2} \big(
  B_{{3}}  \sin  2\,\theta
-  B_{{4}}  \cos  2\,\theta
\big)
 \\&
 +  r\,\big(
   A_{{4}}\,\sin \theta
 +  A_{{3}}\,\cos \theta
+ 9 \, C_1  \,\sin 3\,\theta
+ 9 \, C_2  \,\cos 3\,\theta
 \big)
 \\&
+3\,(  B_{{1}} \,\sin  2\,\theta
-  B_{{2}} \,\cos  2\,\theta
-6\, D_{{1}} \,\sin  4\,\theta
+6\, D_{{2}} \,\cos  4\,\theta)
 \Big] \, R''
 \\[5pt]&
 +3\,{r}^{5}
 \Big[
    2\,{r}^{3}\big(
   A_{{2}} \,\sin  \theta
+  A_{{1}} \,\cos  \theta
\big)
    + 4\,{r}^{2}  \big(
      \, B_{{4}}\cos  2\,\theta
    - \, B_{{3}}\sin  2\,\theta  \big)
 \\&
  + 2\,r\,  \big(
  A_{{4}} \sin \theta
+ A_{{3}}\,\cos  \theta
- 3\, C_1 \,\sin 3\,\theta
- 3\, C_2 \,\cos 3\,\theta
 \big)
 \\&
 -2( B_{{1}}\,\sin  2\,\theta
 - B_{{2}}\,\cos  2\,\theta
 +2 D_{{2}}\,\cos  4\,\theta
 -2 D_{{1}}\,\sin  4\,\theta) \
 \Big]  R^{(3)}  \,.
\label{CCsep}
\end{aligned}
\end{equation}
}

where $\Theta(\theta)$ is an arbitrary function of $\theta$. Since $S(\theta)$
and $R(r)$ are functions of one variable only, (\ref{CCsep}) is no longer a PDE.
We will obtain several ODEs from it. We differentiate (\ref{CCsep}) twice with
respect to $r$. This eliminates $S(\theta)$ and $\Theta(\theta)$ from the
equation. We then expand in a basis of linearly independent trigonometric
functions $ \sin\theta, \sin2\theta, \sin3\theta, \sin4\theta$, $ \cos\theta,
\cos2\theta, \cos3\theta,
\cos4\theta$ and obtain  the following set of 8 equations that $R(r)$ must
satisfy simultaneously:
{\small
\begin{subequations}
\begin{align}
\label{Ra}
&  12 \left(A_3 - 15 \,r^2\, A_{1}\right) R'
-   12 r \left(A_3 + 27\, r^2\, A_{1}\right) R''
-    r^2 \left(39\, A_3 + 146\, r^2\, A_{1}\right) R^{(3)}  \nonumber \\
&-r^3 \left( 13\, A_3 + 22 \,r^2\, A_{1}\right) R^{(4)}
-r^4 \left(A_3+r^2\, A_{1}\right) R^{(5)} \ = \ 0 \ , \\[10pt]
\label{Rb}
& 2\,  \left(9 B_2 - 40\, r^2\, B_{4}\right) R'
- 2\,r \left(9 B_2 - 40\, r^2 \,B_{4}\right) R''
- r^2 \left(B_2 - 128\, r^2\, B_{4}\right) R^{(3)}  \nonumber \\
& + r^3 \left(7\,B_2 + 32\,r^2\, B_{4}\right) R^{(4)}
+ r^4 \left( B_2 + 2 \,r^2\, B_{4}\right) R^{(5)}  \ = \ 0 \ , \\[10pt]
\label{Rc}
& C_2 \left(36 \,R'-36\, r\, R''+3 \,r^2\, R^{(3)}+9 \,r^3\, R^{(4)}+r^4\,
R^{(5)}\right) \ = \ 0 \ ,\\[10pt]
\label{Rd}
& D_2\left(96\, R - 6\, r\, R'  - 42\, r^2\, R'' + 19 \,r^3 \,R^{(3)} - r^4\,
R^{(4)}  - r^5 \,R^{(5)}    \right) \ = \ 0 \ , \\[10pt]
\label{Re}
&  24 \,\left(  A_4  -  15\, r^2 \,A_2\right) R'
-24 r \,\left(  A_4  +  27 \,r^2\, A_2\right) R''
-r^2 \,\left( 78 \,A_4  + 292 \,r^2 \,A_2\right) R^{(3)}\nonumber  \\
& -r^3 \,\left(26 \,A_4 + 44 \,r^2\, A_2\right) R^{(4)}
-2\,r^4 \,\left( A_4 +  r^2\, A_2\right) R^{(5)} \ = \ 0 \ , \\[10pt]
\label{Rf}
& 2\,\left(9 B_1 - 40\, r^2\, B_3\right) R'
- 2\,r \,\left(9 B_1 - 40 \,r^2\, B_3\right) R''
-  r^2 \,\left( B_1  - 128 \,r^2\, B_3\right) R^{(3)} \nonumber  \\&
+ r^3 \,\left(7\,B_1 + 32\,r^2 \,B_3\right) R^{(4)}
+ r^4\, \left(B_1 + 2 \,r^2 \,B_3\right) R^{(5)}   \ = \ 0 \ , \\[10pt]
\label{Rg}
& C_1\left(36\,R'-36\, r\,R''+3 r^2\,R^{(3)}+9 \,r^3\,
R^{(4)}+r^4\,R^{(5)}\right) \ = \ 0 \ , \\[10pt]
\label{Rh}
& D_1\left(96 \,R - 6\, r\,R'  -  42 \,r^2\,R''  +  19\, r^3\,R^{(3)} -
r^4\,R^{(4)}  - r^5\,R^{(5)}  \right) \ = \ 0 \ .
\end{align}
\end{subequations}
}
Taking linear combinations of eqs. (\ref{Ra}-\ref{Rh}), we get the following
Euler-Cauchy type differential equations
\begin{subequations}
\begin{align}
\label{R4a}
& (  A_{1}\,A_4 -  A_2\,A_3)\bigg[ 180\, R'+324 r\, R''+146\, r^2\, R^{(3)}+22\,
r^3 \,R^{(4)}+r^4\, R^{(5)}  \bigg] \ = \  0  \ ,
\\ \label{R4b}
&  (B_{1}\,B_{4} - B_2\,B_3)\bigg[ 40\, R'-40\, r\, R''-64\, r^2\, R^{(3)}-16\,
r^3\, R^{(4)}-r^4\, R^{(5)}  \bigg] \ = \  0  \ ,
\\  \label{R4c}
&  (\, C_1^2 + C_2^2\, )\bigg[ 36\, R'-36 \,r \,R''+3 \,r^2\, R^{(3)}+9\, r^3
\,R^{(4)}+r^4\, R^{(5)} \bigg] \ = \  0  \ ,
\\  \label{R4d}
&  (\,D_1^2+D_2^2 \, )\bigg[ 96\, R-6 \,r\, R'-42\, r^2\, R''+19\, r^3\,
R^{(3)}-r^4\, R^{(4)}-r^5\, R^{(5)} \bigg] \ = \  0  \ .
\end{align}
\end{subequations}

In particular, the above equations have solutions
\begin{equation}
\begin{aligned}
& R(r)\ = \ \frac{\alpha _1}{r^5}+\frac{\alpha _2}{r^4}+\frac{\alpha _3}{
r^2}+\frac{\alpha _4}{r}+\alpha _5 \,,
&  (  A_{1}\,A_4 -  A_2\,A_3) \neq 0 \ ,
\\ & R(r)\ = \  \frac{\alpha _1}{ r^4}+\frac{\alpha _2}{r^3}+\frac{\alpha
_3}{r}+r^2 \alpha _4+\alpha _5  \,,
&  (B_{1}\,B_{4} - B_2\,B_3) \neq 0 \ ,
\\ &  R(r)\ = \ \frac{\alpha _1}{r^3}+\frac{\alpha _2}{r^2}+r^2 \alpha _3+r^4
\alpha _4+\alpha _5  \,,
&   C_1^2 + C_2^2 \neq 0 \ ,
\\ &  R(r)\ = \ \frac{\alpha _1}{r^2}+\frac{\alpha _2}{r}+r^2 \alpha _3+r^4
\alpha _4+r^6 \alpha _5  \,,
&  D_1^2+D_2^2  \neq 0 \ ,
\label{RS4}
\end{aligned}
\end{equation}
respectively. Otherwise, for
\[
 (A_{1}\,A_4 -  A_2\,A_3) \,= \, 0\,,\quad (B_{1}\,B_{4} - B_2\,B_3)\,= \,
0\,,\quad C_1^2 + C_2^2\,= \, 0\,,\quad D_1^2+D_2^2\,= \, 0 \ ,
\]
the equations (\ref{R4a})-(\ref{R4d}) are trivially satisfied with arbitrary
$R(r)$, but then (\ref{Ra})-(\ref{Rh}) are satisfied only when all parameters
$A_i,\,B_i,\,C_i$ and $D_i$, vanish (so that no fourth order integral $Y$
exists).

The compatibility between common solutions of (\ref{Ra})-(\ref{R4d}) and the
determining equations (\ref{Eq30p})-(\ref{Eq01p}), shows that when (\ref{CCsep}) is satisfied trivially then the most general
form of the radial part $R(r)$ of the potential $V(r,\,\theta)$ is
\begin{itemize}
\item R(r) \ = \ $b\,r^2$ ;    \, in this case all parameters are zero except
$B_3,\,B_{4}$.
\item R(r) \ = \ $\frac{a}{r}$ ; \  all parameters are zero except $A_1,
A_2,\,B_3,\,B_{4}$ .
\item R(r) \ = \ 0 ; \  all parameters are zero except  $A_1, A_2,\,B_3,\,B_{4}$
\ .
\end{itemize}
Exotic potentials $V(r,\,\theta)$ with radial part $R(r)=b\,r^2,\,\frac{a}{r}$
or $0$ are also the only ones that could allow a third order
integral~\cite{TremblayW:2010}.

\section{NONCONFINING POTENTIAL  $V(r,\theta)=\frac{S(\theta)}{r^2}$
}
\label{Potentials}

This potential corresponds to $R(r)=0$. In this article we are interested in
exotic potentials. Eq. (\ref{CCsep}) is linear, so it must be
satisfied trivially. Hence all parameters in (\ref{YA}) vanish except $A_1,
A_2,\,B_3,\,B_{4}$. It is worth mentioning that the singular potentials of the form $V(r,\theta)=
\frac{S(\theta)}{r^2}$ require a renormalization scheme in order to obtain
a well defined problem with a discrete spectrum~\cite{Gupta:1993,Camblong:2000,EssinGriffiths2006}.

The equations (\ref{Eq30p}) - (\ref{Eq12p}) corresponding
to the determining equations ${\cal A}_{rrr}={\cal A}_{rr\theta}={\cal
A}_{r\theta\theta}={\cal A}_{\theta\theta\theta}=0$, respectively,  take the
form
{\footnotesize
\begin{subequations}
\begin{align}
\label{Eq30R0}
&{G_1}^{(1,0)}   =0\,,
\\[25pt] 
\label{Eq03R0}
& {\frac {1}{{r}^{2}} \left(
  {G_2}^{(0,1)}  +  {\frac {{G_3}  }{r}} \right) } =
 \nonumber \\&
-\,{\frac {2 }{{r}^{3}}
\left(
 A_{{2}}\,\sin \theta +A_{{1}}\,\cos \theta
+{\frac {2}{r}} \big(
B_{{4}}\,\cos  2\,\theta   -B_{{3}}\,\sin 2\,\theta
\big)  \right) }\, S
\nonumber \\&
+
\left(
{\frac{4}{{r}^{3}}}
\big( A_{{2}}\,\cos \theta  -\,A_{{1}}\,\sin \theta  \big)
-
{\frac{4}{{r}^{4}}}
\big(
B_{{3}}\,\cos  2\,\theta  + B_{{4}}\,\sin 2\,\theta
\big)
 \right)  \, S'    \,,
\\[25pt] 
\label{Eq21R0}
& {G_3}^{(1,0)}
+ {\frac {1}{{r}^{2}}}  {G_1}^{(0,1)}
 =
{\frac {2}{{r}^{2}}}
\left(
B_{{3}}\,\cos  2\,\theta   + B_{{4}}\,\sin  2\,\theta
\right)
\, S'    \,, \\[25pt] 
\label{Eq12R0}
& {\frac {1}{{r}^{2}}} {G_3}^{(0,1)}   + {G_2}^{(1,0)} + \,{\frac
{2}{{r}^{3}}}{G_1}   =
-\,{\frac {4}{{r}^{3}}}
\left(
 B_{{3}}\,\cos  2\,\theta  - B_{{4}}\,\sin 2\,\theta
\right) \, S
\nonumber \\&
+  \frac {3 }{{r}^{2}}
 \left(
A_{{2}}\,\sin \theta
+\,A_{{1}}\,\cos \theta
+ {\frac {2}{r}}
\big(
\,B_{{4}}\,\cos  2\,\theta   - B_{{3}}\,\sin  2\,\theta
\big)
\right)  \,S'  \ .
\end{align}
\end{subequations}
}
In particular, the equations (\ref{Eq30R0}), (\ref{Eq21R0}) and  (\ref{Eq12R0})
 define the $r$ dependence of the functions $G_{1,2,3}$.
Indeed, from (\ref{Eq30R0}) we obtain
\begin{equation}
{G_1}(r, \theta) = \beta_1(\theta)\ .
\label{G1}
\end{equation}
Substituting (\ref{G1}) into Eq. (\ref{Eq21R0}) and integrating we get:
\begin{equation} 
{G_3}(r, \theta) = -  \frac{2}{r}(B_{{3}}\,\cos \,2\theta  +
B_{{4}}\,\sin\,2\theta  )\, S'
+   \frac{1}{r} \beta_1' (\theta)   + \beta_3(\theta)\ .
\end{equation}

Substituting $G_1,G_3$ into Eq (\ref{Eq12R0}), we find
\begin{equation} 
\begin{split}
G_2(r, \theta) &=   \frac{2}{r^2} ( B_{{3}}\,\cos\,2\theta\,    +
B_{{4}}\,\sin\,2\theta\,  ) S
\\&
+ \left[
-\frac{3}{ r}  \Big( \,A_{{1}}\,\cos\,\theta\,\, +  A_{{2}}\,\sin\,\theta\,
\Big)
- \frac{5}{ r^2}  \Big( B_{{4}}\,\cos\,2\theta - B_{{3}}\,\sin\,2\theta\,\Big)
\right]\,S'
\\&
- \frac{1}{r^2}\,( B_{{3}}\,\cos\,2\theta\,+ B_{{4}}\,\sin\,2\theta\,)\, S''
\\&
+\beta_2(\theta) +\frac{1}{r}\, \beta_3' (\theta)
+\frac{1}{2 r^2}\,\Big(    2\beta_1(\theta) +  \beta_1'' (\theta) \Big)\,.
\end{split}
\end{equation}
Let us now determine the functions $\beta_i(\theta)$.
Substituting the above functions $G_1,G_2,G_3$ into (\ref{Eq03R0}), {i.e.} into
the determining equation
${\cal A}_{\theta\theta\theta}=0$, and collecting in powers of $r$ one finds the
following three equations which define the functions $\beta_{1,2,3}$:
\begin{subequations}
\begin{align}  
\beta_2' (\theta) &=0 \,,
\label{Attta}\\[15pt]
    \beta_3''  (\theta)   +\beta_3 (\theta)
& =
 -2\,\big( \,A_{{1}}\,\cos \theta   +  A_{{2}}\,\sin \theta   \big) \, S \,
\nonumber
\\&
+ 7 \Big(  A_{{2}}\,\cos \theta    - \,A_{{1}}\,\sin \theta   \Big)\, S'
\nonumber
\\&
+3\,\big(A_{{1}}\, \cos \theta     +    A_{{2}}\,\sin  \theta    \big)\, S'' \,,
\label{Atttb} \\[15pt]
  \frac{1}{2}\,\beta_1''' \left( \theta \right) +2\,\beta_1' \left( \theta
\right) &=
   \Big( B_{{3}}\,\cos  2\,\theta  + B_{{4}}\,\sin  2\,\theta      \Big) \, S'''
\nonumber\\&
 + 7\,\Big(B_{{4}}\,\cos  2\,\theta  -   B_{{3}}\,\sin  2\,\theta \Big)\,S''
\nonumber \\&
- 14 \, \Big( B_{{3}}\,\cos  2\,\theta  + B_{{4}}\, \sin  2\,\theta    \Big)\,
S'
\nonumber\\&
- 8\, \Big( B_{{4}}\,\cos  2\,\theta  -   B_{{3}}\,\sin  2\,\theta  \Big)\, S
\,.
\label{Atttc}
\end{align}
\end{subequations}

Equation (\ref{Attta}) implies that
\begin{equation}
\label{bet2}
 \beta_{2}(\theta ) = c_{21}\ ,
\end{equation}
where $c_{21}$ is a constant.

\bigskip

Next, replacing
\[S(\theta) = T'(\theta)\ ,\]
into (\ref{Atttc}) and solving this equation we find the function
$\beta_1(\theta)$:
\begin{equation}
\begin{split}
\beta_1(\theta) &=  2\,\left( B_{{4}}\,\cos  2
\,\theta -B_{{3}}\,\sin  2\,\theta   \right) T
+ 2\,\left( B_{{4}}\, \sin 2\,\theta  +B_{{3}}\,\cos  2\,\theta
 \right) \, T'
\\&
+ \frac{1}{2}\,c_{11} \,\sin  2\,\theta
- \frac{1}{2} \,c_{12}\,\cos  2\,\theta
+ c_{13}\ ,
\label{gam1}
\end{split}
\end{equation}
where the $c$'s are integration constants.

Similarly, the solution to equation (\ref{Atttb}) provides the function
$\beta_3(\theta)$
\begin{equation}
\begin{split}
\beta_{{3}} \left( \theta \right) &=
{c_{31}}\,\cos  \theta
+{c_{32}}\,\sin  \theta
+\left( A_{{2}}\,\cos \theta \, - \, A_{{1}}\, \sin  \theta \,
 \right)\, T
\\&
 + 3 \left(\,A_{{2}}\,\sin \theta  +\, A_{{1}}\,\cos  \theta\,  \right)
{T'}\ .
\end{split}
\end{equation}
Now let us turn to the equations (\ref{Eq10p})-(\ref{Eq01p}). From the equation
(\ref{Eq10p}), ${\cal C}_r=0$, we find the function
$G_4(r,\theta)$:
\begin{align}
\label{G4}
 G_4(r,\theta)& =  \Bigg[ -\frac {1}{2\,r} \Big(
 2\,\left( \, A_{{1}}\, \cos  \theta + A_{{2}}\,\sin  \theta \right) S'''
  - 6\, \left(  A_{{2}}\,\cos  \theta
 - A_{{1}}\,\sin  \theta   \right) \, S''    \nonumber
\\&
\hspace{40pt} + 6 \,
\left(  A_{{1}}\,\cos  \theta  +  A_{{2}} \,\sin  \theta \right) S'
+  \, \beta_{{3}}'''
+                   \, \beta_{{3}}'
\Big)   \nonumber
\\&
-
\frac {1}{{r}^{2}} \Big(  \frac{1}{2}\, \left( B_{{4}}\sin  2\,\theta  +B_{{3}}
\cos  2\,\theta   \right)  {S}^{(4)}
- 4 \left( \,B_{{3}}\sin  2\,\theta  - \,B_{{4}}\cos  2\,\theta
   \right) S'''      \nonumber
\\&
 \hspace{40pt}- 10\, \left( \,B_{{3}}\cos  2\,\theta  + \,B_{{4}}\sin 2\,\theta
 \right) S''   - 8\, (  B_{{4}} \,\cos  2\,\theta  -   B_{{3}}\,\sin  2\,\theta
)S'   \nonumber
 \\&
\hspace{50pt}  - \frac{1}{4}\, \beta_{{1}} ^{(4)} -  \beta_{{1}}''
 \Big)
\Bigg] {{\hbar}}^{2}     \nonumber
\\&
+\beta_{{4}}
- \,{\frac{2}{r}\, S'\, \beta_{{3}}  }
 +\frac{1}{r^2} \Big(
 2\,\left(\, B_{{4}}\,\sin  2\,\theta +B_{{3}}\,\cos  2\,\theta  \right){S'}^2
 -S' \,{\beta_{{1}}}'  +4\,S  \, \beta_{{1}}  \Big)\,.
\end{align}
At this point, all eight coefficients
${\cal A}_{rrr}, {\cal A}_{rr\theta}, {\cal A}_{r\theta\theta}, {\cal
A}_{\theta\theta\theta}, {\cal B}_{rr}$,
${\cal B}_{r\theta}, {\cal B}_{\theta\theta}, {\cal C}_{r}$ in
(\ref{commutator}) vanish. In fact, the main
equation to be solved is ${\cal C}_{\theta}=0$, presented in (\ref{Eq01p}).

Substituting $G_1,G_2,G_3,G_4$ into the determining equation
${\cal C}_{\theta}=0$, (\ref{Eq01p}), and collecting powers of $r$ we get three
equations that must be satisfied simultaneously in order for $Y$  in (\ref{Y})
to be an integral of motion:
\begin{equation}\label{Cta}
0\ = \   {\beta_{{4}}'} \,  -4\,c_{{21}} T'' \ ,
\end{equation}

\begin{equation}
\begin{aligned}
\label{Ctb}
0\ =\ &
 \Bigg[
 -4\left(
  \,A_{{1}} \, \cos  \theta
 + \,A_{{2}}\, \sin  \theta
 \right)\,T'
 - 8 \left(
  A_{{1}}\, \sin  \theta
-\,A_{{2}}\,  \cos  \theta
\right) \,T''
 \\&
 + 6\left(
  \,A_{{1}} \, \cos  \theta
 +\,A_{{2}}\, \sin  \theta
 \right) \,T^{(3)}
+\frac{5}{2} \left(
\,A_{{1}}\,  \sin \theta
-\,A_{{2}}\, \cos  \theta
\right) \,T^{(4)}
 \\&
-\frac{1}{2} \left(
A_{{1}}\,  \cos  \theta
+\,A_{{2}}\, \sin  \theta
\right) \,T^{(5)}
\Bigg] {{\hbar}}^{2}
- 12\left(
 A_{{1}}\,  \cos \theta
+\,A_{{2}}\, \sin  \theta
\right)  \,(\, T' \,)^2
  \\&
+ \Big(
24\left(
\,A_{{2}}\, \cos  \theta
-\,A_{{1}}\, \sin  \theta
 \right) \,T''
 +6 \left(
  A_{{1}}\, \cos \theta
 +\,A_{{2}}\, \sin \theta
 \right) \,T^{(3)}
  \\&
 +4\left(
     A_{{1}}\,  \sin  \theta
   -\,  A_{{2}}\,  \cos \theta
   \right)\,T
   -4\left(
     c_{{31}}\,\cos \left( \theta \right)
   +c_{{32}}\,\sin \left( \theta \right)
\right)
\Big) \,T'
 \\&
+6 \left(
 A_{{1}} \, \cos  \theta
 + \,A_{{2}}\, \sin  \theta
\right)  \,(\, T'' \,)^2
  + \left(
  6\left(
    A_{{1}} \, \cos  \theta
   +\, A_{{2}} \, \sin  \theta
  \right)\,T
  \right. \\&\left.
-6\left(
  c_{{31}}\,\sin  \theta
  -c_{{32}}\,\cos \theta
  \right)
  \right) \,T''
  + \left(
2\,\left(
    A_{{2}} \, \cos  \theta
  -A_{{1}}\, \sin  \theta
  \right) \,T
\right.   \\& \left.
+2\, \left(
   c_{{31}}\,\cos \theta
  +c_{{32}}\,\sin  \theta
\right)
  \right) \,T^{(3)}
  \,,
\end{aligned}
\end{equation}

\begin{equation}
\begin{aligned}
\label{Ctc}
0\ = \ &
 \Bigg[
-32 \left( \,B_{{4}}\cos2\,\theta - \,B_{{3}}\sin  2\,\theta   \right)\,T'
 -40 \left(
  \,B_{{4}}\sin  2\,\theta
 +\,B_{{3}}  \cos  2\,\theta
 \right) \,T''
  \\&
 + 20 \left(
   \,B_{{4}}\cos  2\,\theta
 -\,B_{{3}}\sin  2\,\theta   \right) \,T^{(3)}
 + 5\,\left(
   \,B_{{4}}\sin  2\,\theta
 + \,B_{{3}}\cos  2\,\theta
 \right) \,T^{(4)}
 \\&
  -\frac{1}{2} \left(
  \,B_{{4}}\cos  2\,\theta
 -\,B_{{3}}\sin  2\,\theta
  \right) \,T^{(5)}
  \Bigg] {{\hbar}}^{2}
 -48 \left(
   \,B_{{4}}\cos  2\,\theta
  -\,B_{{3}}\sin 2\,\theta
 \right)  \,(\,T'\, )^2
 \\&
 + \Big[
 -48\left(
  \,B_{{4}}\sin  2\,\theta
 +\,B_{{3}}  \cos  2\,\theta
 \right) \,T''
 + 6\left(
  \,B_{{4}}\cos  2\,\theta
 -\,B_{{3}}\sin 2\,\theta
 \right) \,T^{(3)}
 \\&
+ 32 \left(
  \,B_{{4}}\sin  2\,\theta
 +\,B_{{3}}  \cos  2\,\theta
 \right)\,T
 -8\left(
     c_{{11}}\,\cos  2\,\theta
 + c_{{12}}\,\sin  2\,\theta
 \right)
 \Big] \,T'
 \\&
+ 6\left(
  \,B_{{4}}\cos  2\,\theta
-\,B_{{3}}\sin  2\,\theta
\right)  \,(\, T'' \,)^2
 + \left[
 24\left(
  B_{{3}}  \,\sin  2\,\theta
 -  B_{{4}} \,\cos  2\,\theta
 \right) \, T
\right.  \\& \left.
-6\left(
  c_{{11}}\,\sin  2\,\theta
 - c_{{12}}\,\cos  2\,\theta
 \right)
 \right] \,T''
 -\Big[
 4\, \left(
     \,B_{{4}}\sin  2\,\theta
  +\, B_{{3}}\cos  2\,\theta
 \right)\,T
 \\&
 - c_{{11}}\,\cos  2\,\theta
 - c_{{12}}\,\sin  2\,\theta
 \Big]\,T^{(3)}  \,,
\end{aligned}
\end{equation}
At this stage we assume $\hbar \neq 0$. We see that in the classical case
($\hbar\rightarrow 0$) equations (\ref{Ctb}) and (\ref{Ctc}) simplify greatly.
The above non-linear equations (\ref{Ctb}) and (\ref{Ctc}) will determine the
angular part of the potential. They both pass the Painlev\'e test.

Equation (\ref{Cta}) determines the function $\beta_4$
\begin{equation}
 \beta_4(\theta)\ =\ 4\, c_{21}\, S(\theta)   + c_{41}\ ,
\label{b4sp}
\end{equation}
together with (\ref{bet2}), this defines $G_4(r,\,\theta)$ of (\ref{G4})
completely in terms of $S(\theta)$ and some constants.

The parameters $c_{13}$, $c_{21}$ in (\ref{gam1}) and (\ref{b4sp}) can be set
equal to zero by linear combinations of $H$ and $X$. Moreover,  $c_{41}$ in
(\ref{b4sp})
is simply a constant that commutes with $H$ trivially. Therefore, without loss
of generality we choose

\[
c_{13} \ = \ 0 \ , \qquad c_{21} \ = \ 0 \ , \qquad c_{41} \ = \ 0 \ . \qquad
\]

\vspace{0.2cm}

Equations (\ref{Ctb}) and (\ref{Ctc}) depend on mutually exclusive sets of
parameters, namely $(A_1,\,A_2,\,c_{31},\,c_{32})$ and
$(B_3,\,B_4,\,c_{11},\,c_{12})$, respectively. Moreover, for $A_1=A_2=0$
(\ref{Ctb}) reduces to a linear equation, as does (\ref{Ctc}) for $B_3=B_4=0$.
Since we are looking for exotic potentials, all linear equations for $T(\theta)$
must be satisfied identically. Hence we have 2 cases to consider

\begin{itemize}
  \item Case (I)
\begin{equation}
\nonumber
A_1^2+A_{2}^2\neq 0\,,\qquad \ B_3=B_4=c_{11}=c_{12}=0 \ ,
\end{equation}
  By a rotation we can set $A_1=0$\  ,
  \item Case (II)
\begin{equation}
\label{Cases}
B_3^2+B_{4}^2\neq 0\,,\qquad \ A_1=A_2=c_{31}=c_{32}=0 \ ,
\end{equation}
  By a rotation we can set $B_4=0$ \ .

In Case I and II one of the two equations (\ref{Ctb})-(\ref{Ctc}) trivializes,
so only one nonlinear equation must be solved. It already passed the Painlev\'e
test.

  \item Case (III) $A_1^2+A_{2}^2\neq 0\,, \qquad B_3^2+B_{4}^2\neq 0$

  In this case the two nonlinear determining equations (\ref{Ctb}) and
(\ref{Ctc}) remain. Thus they will either be incompatible or $T(\theta)$ will be
a very special case of the solutions obtained in Case I and Case II. We shall
not investigate this case further since it cannot provide any new exotic
potentials.
\end{itemize}

We also note that in the quantum case (\ref{Ctb}) and (\ref{Ctc}) are fifth
order equations. In the classical limit $\hbar\rightarrow 0$ they reduce to
third order ones, to be considered in Section 6.

\subsection{Case I, $A_2\neq0$, $A_1=B_3=B_4=0$}

Equation (\ref{Ctc}) is linear and must be satisfied trivially, so we have
$c_{11}=c_{12}=0$.
Equation (\ref{Ctb}) simplifies to
\begin{align}  
&A_{2}\,\Bigg[  2 \,\sin \theta \, T'
- 4 \,\cos  \theta   \, T''
- 3 \,\sin  \theta   \,  T'''
+ \frac{5}{4} \,\cos \theta   \, T^{(4)}
+  \frac{1}{4} \,\sin  \theta    \, T^{(5)}
\Bigg]\, {{\hbar}}^{2}
\nonumber \\&
+ A_{2}\,\Big( 2 \,\cos  \theta   \,  T'
+ 3  \,\sin  \theta   \, T''
-  \cos  \theta   \, T'''
 \Big) T  + 6  \,\sin  \theta A_{2}  \, T'^{2}
\nonumber \\&
  - \Big(  12\,A_{2} \,\cos \theta  \, T''
+ 3  \,\sin  \theta  A_{2}   \,T'''
-2\,( c_{31}\,\cos  \theta  +  c_{32}\,\sin  \theta  ) \Big)  \,T'
- 3\, A_{2}  \,\sin  \theta     \left( T''\right)^{2}
\nonumber \\&
- 3\,\left(\, c_{32}\,\cos \theta  - c_{31}\,\sin
\theta \,  \right) \, T''
-  \left( c_{31}\,\cos  \theta  + c_{32}\,\sin  \theta  \right)  \,T'''
=0\ .
\label{Tnle5theta}
\end{align}
This equation can be integrated once resulting in the 4-th order equation
\begin{align}
 & A_{2}\,\Bigg[  2 \,\cos \theta\,  T'
+ 2 \,\sin  \theta \,  T''
-  \cos \theta  \,T'''
-\frac{1}{4}  \,\sin
\theta  \,T^{(4)}
 \Bigg] {{\hbar}}^{2}
\nonumber  \\&
 +
 A_{2}\,\Big(  \cos  \theta  \,T''  -2 \,\sin  \theta  \, T'
\Big) T
+ 4\,A_{2}\,\cos  \theta \,
\left( T'\right) ^{2}
\nonumber  \\&
 + \Big(  3\, A_{2} \,\sin  \theta \,  T''
+2\, (\,c_{32}\,\cos  \theta  - c_{31}\,\sin  \theta \, )
 \Big) \,T'
+
 \left( c_{31}\,\cos  \theta  +  c_{32}\,\sin  \theta
  \right) \,T''
 + \frac{1}{2}\, K_1 = 0 \,,
\nonumber
\end{align}
where $K_1$ is an arbitrary integration constant.
Transforming to the variable

\[
z\ =\ \tan\,\theta \ ,
\]
and dividing by $(1/4) (z^2+1)^2$, we get:

\begin{align}
&A_{2}\,\Bigg[ 24\,
{z}^{2}\, T' + 12\, \left( 3\,{z}^{2}+2\,
 \right) z \, T''
 +4\, \left( {z}^{2}+1 \right)  \left( 3\,{z}^{2}+ 1 \right) \, T'''
+ \left( {z}^{2}+1 \right) ^{2} \, z \, T^{(4)}
 \Bigg]\, {{\hbar}}^{2}
\nonumber\\&
-  4\,A_{{2}} \,T''\, T
- A_{2}\,\left( 24\,{z}^{2}+16\, \right)  \left( T'
\right) ^{2}
- \left( 12\, \left( {z}^{2}+1 \right) \,z\,A_{{2}} \, T''
+8\,c_{32} \right) \,T'
\nonumber \\&
- 4\left( c_{32}\,z + \,c_{31} \right) \,T''
- 2\,{\frac {{ K_1}}{ \left( {z}^{2}+1 \right) ^{3/2}}} = 0 \ .
\end{align}

Putting ${ c}_{31}\rightarrow  2\,A_2\,{{c}_{31}},\, { c}_{32}\rightarrow
2\,A_2\,{{c}_{32}} , \, \ {{K}_1} \rightarrow  2\,A_2\,K_1$, we integrate the
above equation, using $z$ as integrating factor to get
the following third order non-linear differential equation
\begin{align}
&  \Big[
 2\left( 1 -3\,{z}^{4}  \right) \,T'
-2\,z \left( 3\,{z}^{2}+1 \right)  \left( {z}^{2}+1 \right) \,T''
 -{z}^{2} \left( {z}^{2}+1 \right)^{2} \, T'''
\Big]\, {{\hbar}}^{2}
\nonumber \\&
-2\,T ^{2}+ 4\left( \,z\, T' - 2\,{c}_{31}\right) T
+ 6\,{z}^{2} \left( {z}^{2}+1 \right)
 \left( T'\right) ^{2}
+ 8\,z \left( {c}_{32}\,z + {c}_{31} \right) \,T'
\nonumber \\&
 -4\,{\frac {{{K}_1}}{\sqrt {{z}^{2}+1}}}  +  {K_2}  =0\,,
\label{NLE1}
\end{align}
here $K_2$ is another arbitrary integration constant. The transformation
$(z,T(z))\mapsto(x,W(x))$:
\begin{equation}
 z=\frac{2\,\sqrt{x}\,\sqrt{1-x}}{1-2\,x},\quad
T=\frac{ \hbar^2\, W}
{\sqrt{x}\sqrt{1-x}}
+\frac{ (3\,\hbar^2 + 8\, {c}_{32})\,(1-2\,x)}
{ 8\, \sqrt{x}\,\sqrt{1-x}}
- 2\,{c}_{31}   \ ,
\label{transf}
\end{equation}
 maps (\ref{NLE1})  to an equation  contained in a series of papers by C.
Cosgrove (see for
example~\hbox{[\!\!\citenum{CosgroveIX-XI,Cosgrove:1993,Cosgrove:2006,
Cosgrove:2006}]} )
 on higher order Painlev\'e equations. Equation (\ref{NLE1}) is mapped into the
third order
differential equation Chazy-I.a with parameters
\begin{equation}
\begin{split}
\label{ccc}
q_1=q_4=q_5=q_6 =0,\quad
q_2=-q_3=1,\quad
q_7=\frac{ 5\, \hbar^2 + 16 \,{c}_{32}}{16\,\hbar^2},   \\
q_8= \frac{{K}_1}{\hbar^4}, \quad
q_9=-\frac{32\, {{K}_1} + 8\,{K}_2 + 3\,\hbar^4  + 64 \,{{c}_{31}^2}
+ 32\,\hbar^2\,{{c}_{32}} + 64\, {{c}_{32}}^2}{64\,\hbar^4}\,.
\end{split}
\end{equation}

The equation for $W(x)$ can be integrated, and the resulting
non-linear second order differential equation becomes the equation
SD-I.a in Cosgrove's paper~\cite{CosgroveIX-XI}
\begin{align}
 {( W'')}^2 \  &= \ -\frac{4}{f^2(x)} \bigg[q_1{(x\,W'-W)}^3  +q_2
W'\,{(x\,W'-W)}^2
\label{Sec} \\&
 +  q_3 {( W')}^2\,{(x\,W'-W)} + q_4 {( W')}^3  +  q_5{(x\,W'-W)}^2
\nonumber \\&
+  q_6 W'\,{(x\,W'-W)}  + q_7 {( W')}^2  + q_8{(x\,W'-W)}+ q_9\,W' + q_{10}
\bigg]\ ,
\nonumber
\end{align}

The integration constant $q_{10}$ is arbitrary and the function $f(x)$ satisfies
$f(x)=q_1\,x^3+q_2\,x^2+q_3\,x+q_4$. Eq. (\ref{Sec}) is the first canonical
subcase of the more general equation that
Cosgrove called the ``master Painlev\'e equation''. Equation SD-I.a is solved by
the Backlund correspondence
\begin{align}
 W(x)&=\frac{x^2(x-1)^2}{4P_6(P_6-1)(P_6-x)}\bigg[P_6'-\frac{P_6(P_6-1)}{x(x-1)}
\bigg]^2+\frac{1}{8}(1-\sqrt{2\gamma_1})^2(1-2P_6)\nonumber\\
&-\frac{1}{4}\gamma_2\bigg(1-\frac{2x}{P_6}\bigg)-\frac{1}{4}
\gamma_3\bigg(1-\frac{2(x-1)}{P_6-1}\bigg)+\bigg(\frac{1}{8}-\frac{\gamma_4}{4}
\bigg)\bigg(1-\frac{2x(P_6-1)}{P_6-x}\bigg) \ ,
\label{Wpot}
\end{align}
and
\begin{align}
 W'(x)=-\frac{x(x-1)}{4P_6(P_6-1)}\bigg[P_6'-\sqrt{2\gamma_1}\frac{P_6(P_6-1)}{
x(x-1)}\bigg]^2-\frac{\gamma_2(P_6-x)}{2(x-1)P_6}-\frac{\gamma_3(P_6-x)}{
2x(P_6-1)} \ ,
\label{Wpotd}
\end{align}
where $\sqrt{2\gamma_1}$ can take either sign and  $\gamma_1,\gamma_2,\gamma_3$
and $\gamma_4$ are the arbitrary parameters that define the sixth Painlev\'e
transcendent $P_6$ which satisfies the well known second order differential
equation:
\begin{eqnarray}
\label{P6}
 P_6''=\frac{1}{2}\bigg[\frac{1}{P_6}+\frac{1}{P_6-1}+\frac{1}{P_6-x}\bigg]
(P_6')^2-\bigg[\frac{1}{x}+\frac{1}{x-1}+\frac{1}{P_6-x}\bigg]P_6'\nonumber\\
+\frac{P_6(P_6-1)(P_6-x)}{x^2(x-1)^2}\bigg[\gamma_1+\frac{\gamma_2\,x}{P_6^2}
+\frac{\gamma_3\,(x-1)}{(P_6-1)^2}+\frac{\gamma_4\,x(x-1)}{(P_6-x)^2}\bigg] \ .
\end{eqnarray}

Thus, we have
\[
W \ =\ W(x\,;\,\gamma_1,\,\gamma_2,\,\gamma_3,\,\gamma_4)  \ .
\]

The parameters $\gamma_1,\gamma_2,\gamma_3$ and $\gamma_4$ are related to the
arbitrary constants of integration ${c}_{31}, {c}_{32}, {K}_1$ and $K_2$ through
the relations
\begin{align}
&-4q_7= \gamma_1-\gamma_2+\gamma_3-\gamma_4-\sqrt{2\gamma_1}+1,  \nonumber  \\
&-4q_{8}=(\gamma_2+\gamma_3)(\gamma_1+\gamma_4-\sqrt{2\gamma_1}),   \nonumber
\\
&-4q_9=(\gamma_3-\gamma_2)(\gamma_1-\gamma_4-\sqrt{2\gamma_1}+1)+\frac{1}{4}
(\gamma_1-\gamma_2-\gamma_3+\gamma_4-\sqrt{2\gamma_1})^2, \nonumber \\
&-4q_{10}=\frac{1}{4}(\gamma_3-\gamma_2)(\gamma_1+\gamma_4-\sqrt{2\gamma_1}
)^2+\frac{1}{4}(\gamma_2+\gamma_3)^2(\gamma_1-\gamma_4-\sqrt{2\gamma_1}+1)
\label{ParcA}
\end{align}
In particular, (\ref{ccc}) together with (\ref{ParcA}) imply that the constants
$c_{{31}}$ and $c_{{32}}$ can be written in terms of the $\gamma$'s.

A superintegrable potential expressed in terms of the Painlev\'e transcendent
$P_6$ was obtained earlier~\cite{TremblayW:2010}. It allowed a third order integral
and required a specific relation between the constants $\gamma_1,...,\gamma_4$.
Here we obtain the most general form of $P_6$.

From the inverse transformation $x\to z=\tan\theta$ in (\ref{transf}) we get
\begin{equation}
x_{\pm}=\frac{1}{2}\pm\frac{1}{2\sqrt{1+z^2}}  =\Bigg\{\begin{array}{ll}
    \sin^2\big(\frac{\theta}{2}\big)\\\\
    \cos^2\big(\frac{\theta}{2}\big)\\
    \end{array}
\end{equation}

we obtain two solutions for $S(\theta)=T'(\theta)$. For the Case I we obtain two
quantum potentials

\begin{equation}
\begin{aligned}
\label{VcaseI}
 V(r,\theta)& = \  \frac{\partial_{\theta}T(x_{\pm})}{r^2} \\ &  =
\frac{\hbar^2}{r^2}\Bigg(\,W'(x_{\pm}) \mp  \frac{2\,\cos \theta}{\sin^2\theta }
 \,W(x_{\pm})
+ \frac{1}{2\,\sin^2\theta }\,\Gamma  \Bigg) \ ,
\end{aligned}
\end{equation}

where $\Gamma=(\gamma_2+\gamma_4+\sqrt{2\,\gamma_1}-\gamma_1-\gamma_3-\frac{3}{8})$.
Both $T$ and $W$ are completely defined through (\ref{transf})-(\ref{ParcA}).
The integral $Y$ in both cases is
\begin{equation}
 \begin{split}
 \label{YcaseI}
 Y \ =& \    \hbar^4\,\{ \partial_\theta^3,\, \sin\theta\, \partial_r \}\,
+  \frac{\hbar^4}{r}\{ \partial_\theta^3,\,  \cos\theta \partial_\theta\}  -
 \ \hbar^2\,\{ G_1(r,\theta),\,\partial_r^2  \} \ - \ \hbar^2\, \{ G_3(r,\theta),\,\partial_r\,\partial_\theta  \}
\\&
\ - \  \hbar^2\,\{ G_2(r,\theta),\,\partial_\theta^2  \}
\ + \  G_4(r,\theta)\  ,
\end{split}
\end{equation}
 ($A_2=1$) where
\begin{align}
 &   G_{{1}} \left( r,\theta \right) =0  \,, \\
 &   G_{{2}} \left( r,\theta \right)  =  \frac{1}{r} \,
\Big(
4\,\cos  \theta \ T'
 +2\,c_{{32}}\,\cos  \theta
- \left( T + 2\,c_{{31}}   \right) \sin  \theta
\Big)  \,, \nonumber \\
&
 G_{{3}} \left( r,\theta \right) = 3\,\sin  \theta \ T'   +
 \left( T   +2\,c_{{31}}\, \right) \cos  \theta
  +2\,c_{{32}}\,\sin \theta  \ ,\nonumber \\
&
G_{{4}} \left( r,\theta \right) =  \,
{\frac {1}{2{r}}} \Big( \,\sin \theta \ T^{(4)}
+ 4\,\cos \theta  \  T^{(3)}
- 3\,\sin  \theta  \ T''
- 2\,\cos  \theta  \ T'
\Big) {{\hbar}}^{2}
\, \nonumber \\ &
-
{\frac {2}{{r}}} \Big(
  3\,\sin  \theta  \ T'
+ \,\cos  \theta  \ T
+ 2\,c_{{31}}  \cos  \theta
+ 2\,c_{{32}}  \sin  \theta     \Big) \,T''
 \ .\nonumber
\end{align}
here $T'=\partial_{\theta}T(x_{\pm})$. The integral $Y$ and the corresponding
potential $V(r,\theta)$ depend on the same constants, namely, the four
parameters $\gamma_1,\gamma_2,\gamma_3,\gamma_4$ in (\ref{ParcA}) which define
the sixth Painlev\'e transcendent $P_6$.

\bigskip

\subsection{Case II, $B_3\neq 0$,\ $A_1=A_2=B_4=0$}

Equation (\ref{Ctb}) reduces to a linear one that must be satisfied trivially so we have to impose $c_{31}=c_{32}=0$.
Equation (\ref{Ctc}) simplifies to
\begin{align}
& B_3\,\Bigg[ 16 \, \sin 2\,\theta  \,  T'
-20 \,\cos 2\,\theta \,  T''
-10 \,\sin  2\,\theta  \, T'''
+\frac{5}{2}  \,\cos  2\,\theta  \,  T^{(4)}
\nonumber \\&
+ \frac{1}{4}
\,\sin  2\,\theta \,  T^{(5)}
 \Bigg]\, {{\hbar}}^{2}
+ B_3\,\Big(  16
\,\cos 2\,\theta \,
   T'
 + 12  \,\sin  2\,\theta \,  T''
-2   \,\cos 2\,\theta  \,  T'''
 \Big)\, T
\nonumber \\&
   + 24 \,B_{{3}}\,\sin  2\,
\theta \,
 \left(T' \right) ^{2}
- \Big(  24
   \,B_{{3}}\cos  2\,\theta \, T''
+ 3 \,B_{{3}}\sin  2\,\theta \, T'''
+ 4\,(\,c_{11}\,\cos  2\,\theta  +  c_{12}\,\sin  2\,\theta \, ) \Big)\, T'
\nonumber \\&
- 3\,B_{{3}}\,\sin
 2\,\theta \,
  \left( T''  \right)^{2} +
3\, \left( \,c_{12}\,\cos  2\,\theta  - c_{11}\,\sin 2\,\theta\, \right)\, T''
+ \frac{1}{2}\left(
 c_{11}\,\cos 2\,\theta
+c_{12}\,\sin  2\,\theta
 \right)\, T'''  =0 \ ,
\label{Tnle52theta}
\end{align}
This equation  can be integrated once resulting in
\begin{align}
 & B_3\,\Big[
\frac{1}{4}
 \sin  2\,\theta
 \,T^{(4)}
+ 2\, \cos  2\,\theta  \,T^{(3)}
- 6\, \sin  2\,\theta  \,T''
-8\, \cos  2\,\theta \, T'
\Big]\, {{\hbar}}^{2}
\nonumber \\&
+ B_3\,\Big(
8 \,\sin  2\,\theta  \, T'
-2
\,\cos  2\,\theta  \, T''
\Big)\, T
-8
\,B_{{3}}\cos  2\,\theta
\left( T'  \right) ^{2}
\nonumber \\&
+\Big(
2\,\left(\,c_{12}\,\cos  2\,\theta -  c_{11}\,\sin  2\,\theta \, \right) - 3
\,B_{{3}}\,\sin  2\,\theta  \,T''
\Big) \,T'
 + \frac{1}{2}\left( c_{11}\,\cos 2\,\theta + c_{12}\,\sin 2\,\theta \,  \right)
\, T''  + \frac{1}{2}{K_1}
  =0\ .
\end{align}

Putting $z\ =\ \tan 2\theta$, (and dividing by the common factor \hbox{$4 (z^2+1)^{3/2}$}) we obtain

\begin{align}
&  B_{{3}} \,\Big( 4\,\left( 6\,{z}^{2}+1 \right) \, T'
  + \left( 36\,{z}^{3}+ 26\,z  \right) \, T''
+4\, \left( {z}^{2}+1 \right)  \left( 3\,{z}^{2}+1 \right) \, T^{(3)}
 + \left( {z}^{2}+1 \right)^{2} \,z  \, T^{(4)}
 \Big) {{\hbar}}^{2}
\nonumber\\&
  -2  \,B_{{3}}  \, T'' \, T
-B_{{3}} \, \left( 12\,{z}^{2} + 8 \right)  \left( T'  \right)^{2}
+ \left(c_{12} -6\, \left( {z}^{2}+1 \right) \, z\,B_{{3}} \, T''
 \right) \,T'
\nonumber\\&
+ \frac{1}{2}\left( c_{12}\,z+ c_{11} \right) \, T''
+  \frac{1}{8} \,\frac{ {K_1} }{ \left( {z}^{2} + 1 \right)^{3/2} }   =0\ .
\label{NLE2intz}
 \end{align}

We introduce ${c}_{11} \rightarrow 2\,B_{3}\,{c}_{11}$, ${c}_{12} \rightarrow
2\,B_{3}\,{c}_{12}$ and ${K}_{1} = 2\,B_{3}\,{K}_{1}$ and again integrate
(\ref{NLE2intz}) to obtain
\begin{align}
& \Big(  2\left( 3\,{z}^{4} + \,{z}^{2} - 1 \right) \,T'
+2\,z \left( 3\,{z}^{2}+1 \right)  \left( {z}^{2}+1 \right) \, T''
+{z}^{2} \left( {z}^{2}+1 \right) ^{2} \, T^{(3)}
\Big) {{\hbar}}^{2}
\nonumber \\&
+  T^{2}  - \left(  {c}_{11} + 2\,z\, T'   \right) T
 -3\,{z}^{2} \left( {z}^{2}+1 \right)  \left( T' \right)^{2}
+   z \left( {c}_{12} \,z + {c}_{11} \right) \, T'
\nonumber \\&
- \frac{1}{4}\,\frac {{K_1}}{\sqrt {{z}^{2}+1}}    +{K_2} =0 \ .
\label{NLE2}
\end{align}

The transformation $(z,T(z))\mapsto(x,W(x))$:
\begin{equation}
 z=\frac{2\sqrt{x}\sqrt{1-x}}{1-2\,x},\quad
T=\frac{\frac{1}{4} \,(1-2 \,x)\, \left(\hbar ^2-  {c}_{12}\right)+2 \,\hbar^2
\,W(x)}{\sqrt{x \,(1-x)}}+\frac{{c}_{11}}{2}\ ,
\label{transfII}
\end{equation}
 maps (\ref{NLE2})  to an equation  contained in the series of papers by C.
Cosgrove
 on higher order Painlev\'e equations~\cite{CosgroveIX-XI}. Equation
(\ref{NLE2}) is mapped into the third order
differential equation Chazy-I.a  with parameters
\begin{gather}
q_1=q_4=q_5=q_6 =0,\quad
q_2=-q_3=1,\quad
q_7=   \frac{1}{16}-\frac{{c}_{12}}{8\, \hbar ^2}         , \nonumber \\
q_8= -\frac{K_1}{32 \,\hbar ^4}, \quad
q_9= -\frac{{c}_{11}^2+{c}_{12}^2-{K}_1-4 K_2-\hbar ^4}{64\, \hbar ^4}    \ .
\label{ParcB}
\end{gather}

The solution for the function $W(x)$ is given in (\ref{Wpot}), however the
independent variable is different, namely:
\begin{equation}
x_{\pm}=\frac{1}{2}\pm\frac{1}{2\sqrt{1+z^2}}  =\Bigg\{\begin{array}{ll}
    \sin^2\theta\\\\
    \cos^2\theta\ . \\
    \end{array}
\end{equation}
We obtain two solutions for $S(\theta)=T'(\theta)$. By taking the derivative
$\partial_{\theta}T(x_{\pm})$ we obtain the quantum potentials

\vspace{0.2cm}

\begin{equation}
\begin{aligned}
 V(r,\theta)& = \  \frac{\partial_{\theta}T(x_{\pm})}{r^2} \\ &  =
\frac{\hbar^2}{r^2}\Bigg(\,4\,W'(x_{\pm}) \mp  \frac{8\,\cos 2\theta}{\sin^2
2\theta }  \,W(x_{\pm})
+ \frac{1}{\sin^2 2\theta }\,\Gamma  \Bigg) \ ,
\label{VWA7}
\end{aligned}
\end{equation}

where
$\Gamma=2(\gamma_2+\gamma_4+\sqrt{2\,\gamma_1}-\gamma_1-\gamma_3+\frac{3}{4})$,
$T$ is now defined through (\ref{transfII})-(\ref{ParcB}). These potentials
correspond to the integral

\vspace{0.2cm}

{\footnotesize
\begin{align}
 Y  =&\
   \hbar^4\,\big\{ \partial_\theta^2,\,   \cos 2\theta  \big\}\,\partial_r^2  -    \frac{\hbar^4}{r^2}\big\{
\partial_\theta^2,\,  \cos 2\theta \, \partial_\theta^2    \big\}   -
\frac{2\,\hbar^4}{r}\,\big\{ \partial_\theta^2,\,  \sin 2\theta\, \partial_\theta   \big\}\,\partial_r  +
\frac{2 \,\hbar^4}{r^2}\big\{ \partial_\theta^2,\,    \, \sin 2\theta \partial_\theta
  \big\}   -   \frac{\hbar^4}{r}\big\{ \partial_\theta^2,\, {\cos 2\theta}
\big\}\,\partial_r
\nonumber \\&
 - \ \hbar^2\,(\{ G_1(r,\theta),\,\partial_r^2  \} \ + \  \{ G_3(r,\theta),\,\partial_r\,\partial_\theta  \}
\ + \  \{ G_2(r,\theta),\,\partial_\theta^2  \})
\ + \  G_4(r,\theta)\ ,
\label{YWA7}
\end{align}
}
(with $B_{3}=1$) where
\begin{align*}
&
  G_{{1}} \left( r,\theta \right) =
2\,\cos  2\,\theta \ T'
- 2\,\sin  2\,\theta \ T
+  c_{{11}} \,\sin  2\,\theta
- c_{{12}} \,\cos  2\,\theta
\nonumber \,, \\ &
G_{{2}} \left( r,\theta \right) = {\frac {1}{{r}^{2}}} \Big(
  2\,\sin  2\,\theta \ T
- 4\,\cos  2\,\theta \ T'
-
  c_{{11}} \,\sin  2\,\theta
+ c_{{12}} \,\cos  2\,\theta
   \Big) \,,
\nonumber  \\ &
  G_{{3}} \left( r,\theta \right) = \frac{1}{r}
  \Big(    2\,c_{{11}}\,\cos  2\,\theta     - 4\,\cos  2\,\theta \  T
  - 6\,\sin  2\,\theta \  T'
+  2\,c_{{12}}\,\sin  2\,\theta
\Big)
\nonumber \,, \\ &
  G_{{4}} \left( r,\theta \right) =
\frac {1}{{r}^{2}}\, \Big(
\big(
- \frac{1}{2}\,\sin  2\,\theta  \  T^{(4)}
-4\,\cos  2\,\theta \  T^{(3)}
 +10\,\sin  2\,\theta \ T''
+8\, \cos  2\,\theta \ T'
 \big) {{\hbar}}^{2}
 \nonumber \\&
 + \, \left(
    \left( 4\,T   - 2\,c_{{11}} \right) \cos  2\,\theta
 + \left( 6\,T' -  2\,c_{{12}} \right) \sin  2\,\theta
     \right) \,T''
 + \,\left( 8\,T'   - 4\,c_{{12}} \right) \,T'   \cos  2\,\theta
 \nonumber \\&
 - \, \left( 8\,T   - 4\,c_{{11}} \right) \,T' \sin  2\,\theta
\Big)
  \ .
\end{align*}

\section{CONFINING POTENTIALS}
\label{CONFINING POTENTIALS}

\subsection{POTENTIAL $V(r,\theta)= b\,r^2 + \frac{S(\theta)}{r^2}$ }

In this case the compatibility condition (\ref{CCsep}) is satisfied trivially if
all parameters are zero except $B_3,\,B_{4}$. Since we can rotate between these
two terms we set $B_{4}=0.$ The equation for $S(\theta)$ corresponds to Case II
of section \ref{Potentials}.

The only determining equation to solve is (\ref{Tnle52theta}) and the solution
for the function $S(\theta)$ will be the same as for $R(r)=0$. The only difference with the case $R(r)=0$ is reflected in the $G$ functions (\ref{Gfunctionsg}) which does not modify the form of the determining equation (\ref{Tnle52theta}). The corresponding
quantum potentials are

\begin{equation}
\begin{aligned}
 V(r,\theta)& = \ b\,r^2 \ + \ \frac{\partial_{\theta}T(x_{\pm})}{r^2} \\ &  = \
b\,r^2 \ + \  \frac{\hbar^2}{r^2}\Bigg(\,4\,W'(x_{\pm}) \mp  \frac{8\,\cos
2\theta}{\sin^2 2\theta }  \,W(x_{\pm})
+ \frac{1}{\sin^2 2\theta }\,\Gamma  \Bigg) \ ,
\end{aligned}
\end{equation}

where
$\Gamma=2(\gamma_2+\gamma_4+\sqrt{2\,\gamma_1}-\gamma_1-\gamma_3+\frac{3}{4})$.
The function $T$ is defined through (\ref{transfII})-(\ref{ParcB}).

\vspace{0.2cm}

The integral of motion in this case is
{\footnotesize
\begin{equation}
\begin{aligned}
 Y  =&
   \hbar^4\,\big\{ \partial_\theta^2,\,   \cos 2\theta  \big\}\,\partial_r^2  -    \frac{\hbar^4}{r^2}\big\{
\partial_\theta^2,\,  \cos 2\theta \, \partial_\theta^2    \big\}   -
\frac{2\,\hbar^4}{r}\,\big\{ \partial_\theta^2,\,  \sin 2\theta\, \partial_\theta   \big\}\,\partial_r  +
\frac{2\,\hbar^4}{r^2}\big\{ \partial_\theta^2,\,    \, \sin 2\theta \partial_\theta
  \big\}   -   \frac{\hbar^4}{r}\big\{ \partial_\theta^2,\, {\cos 2\theta}
\big\}\,\partial_r
 \\&
 - \ \hbar^2\,(\{ G_1(r,\theta),\,\partial_r^2  \} \ + \  \{ G_3(r,\theta),\,\partial_r\,\partial_\theta  \}
\ + \  \{ G_2(r,\theta),\,\partial_\theta^2  \})
\ + \  G_4(r,\theta)\ ,
\end{aligned}
\end{equation}
}
where
\begin{align}
&
  G_{{1}} \left( r,\theta \right) =
2\,\cos  2\,\theta \  T'
- 2\,\sin  2\,\theta \  T
+  c_{{11}} \,\sin  2\,\theta
- c_{{12}} \,\cos  2\,\theta
\nonumber \,, \\ &
G_{{2}} \left( r,\theta \right) = {\frac {1}{{r}^{2}}} \Big(
  2\,\sin  2\,\theta \ T
- 4\,\cos  2\,\theta \ T'
-
  c_{{11}} \,\sin  2\,\theta
+ c_{{12}} \,\cos  2\,\theta
  \Big)  + 2\,b\,\cos  2\,\theta \, {r}^{2}   \,,
\nonumber  \\ &
  G_{{3}} \left( r,\theta \right) = \frac{1}{r}
  \Big(    2\,c_{{11}}\,\cos  2\,\theta     - 4\,\cos  2\,\theta \  T
  - 6\,\sin  2\,\theta \  T'
+  2\, c_{{12}}\,\sin  2\,\theta
\Big)
\nonumber \,, \\ &
  G_{{4}} \left( r,\theta \right) =
\frac {1}{{r}^{2}}\, \Big(
\big(
- \frac{1}{2}\,\sin  2\,\theta  \, T^{(4)}
-4\,\cos  2\,\theta \  T^{(3)}
 +10\,\sin  2\,\theta \ T''
+8\, \cos  2\,\theta \ T'
 \big) {{\hbar}}^{2}
 \nonumber \\&
 + \, \left(
    \left( 4\,T   - 2\,c_{{11}} \right) \cos  2\,\theta
 + \left( 6\,T' -  2\,c_{{12}} \right) \sin  2\,\theta
     \right) \,T''
 + \,\left( 8\,T'   - 4\,c_{{12}} \right) \,T'   \cos  2\,\theta
 \nonumber \\&
 - \, \left( 8\,T   - 4\,c_{{11}} \right) \,T' \sin  2\,\theta
 \Big)
\nonumber \\&
+
b\,\big[
 -8\,\sin 2\,\theta \  T
+  8\,\cos  2\,\theta \   T'
 +4\, \left( c_{{11}}\,\sin  2\,\theta
           - c_{{12}}\,\cos  2\,\theta
 \right) \big]\, {r}^{2}
\nonumber \\&
   -8\,\cos  2\,\theta \, b\,{{\hbar}}^{2}\,{r}^{2}  \ .
\end{align}

\subsection{POTENTIAL OF THE FORM $V(r,\theta)= \frac{a}{r} +
\frac{S(\theta)}{r^2}$ }
In this case the compatibility condition (\ref{CCsep}) is satisfied trivially if
all parameters are zero except
$A_1, A_2,B_{3}, B_{4}$ (as in the Case of $R(r)=0$).

From the condition $[H,Y]=0$ we obtain two 5-th order non-linear equations
equations in $T(\theta)$ that must be satisfied simultaneously, namely eq.
(\ref{Ctc}) and

\begin{align}
0 &=
 \Bigg[  -4\left(  A_{{1}}\,\cos \,\theta + \,A_{{2}}\,\sin\,\theta  \right)
\,T'
+ 8\,\left( A_{{2}}\,\cos \,\theta -\,A_{{1}}\,\sin\,\theta  \right) \,T''
\nonumber \\&
+ 6\,\left( \,A_{{1}}\,\cos \,\theta +\,A_{{2}}\,\sin \,\theta  \right)
 \,T^{(3)}
+ \frac{5}{2}\,\left(  \,A_{{1}}\,\sin \,\theta - \,A_{{2}}\,\cos
\,\theta\right) \,T^{(4)}
\nonumber \\&
- \frac{1}{2}\,\left( \,A_{{1}}\,\cos \,\theta +\,A_{{2}}\,\sin \,\theta
\right)
\,T^{(5)}
+ 15\,a \left( B_{{3}}\sin \,2\theta  -B_{{4}}\cos \,2\theta  \right)  \Bigg]
\,{\hbar }^{2}
\nonumber\\&
+ \Big[  4\,\left( \,A_{{1}}\,\sin \,\theta-\,A_{{2}}\,\cos\,\theta  \right)
\,T'
-6\, \left( A_{{1}}\,\cos \,\theta +\,A_{{2}}\,\sin \,\theta  \right) \,T''
\nonumber \\&
+ 2\,\left( A_{{2}}\,\cos \,\theta -\,A_{{1}}\,
\sin \,\theta  \right) \,T^{(3)}
+24\,a \left( B_{{4}}\,\sin \,2\theta  + B_{{3}}\,\cos \,2\theta
 \right)  \Big]\, T
\nonumber \\&
 -12 \left( \,A_{{1}}\,\cos \,\theta + \,A_{{2}}\,\sin \,\theta  \right)\,
 \left( T' \right)^{2}
+ \Big[  24\,\left( A_{{2}}\,\cos \,\theta -\,A_{{1}}\,\sin \,\theta \right)
\,T''
\nonumber \\&
+ 6\,\left( \,A_{{2}}\,\sin \,\theta+\,A_{{1}}\,\cos \,\theta \, \right)
\,T^{(3)}
+44\,a\,(B_{{3}}\,\sin \,2\theta  -B_{{4}}\,\cos \,2\theta)
\nonumber\\&
-4\,c_{{31}}\,\cos \,\theta -4\,c_{{32}}\,\sin
 \,\theta  \Big] \,T'
+ 6\,\left( \,A_{{2}}\,\sin \,\theta + \,A_{{1}}\,\cos \,\theta  \right)
\left( T''\right) ^{2}
\nonumber \\&
-6\, \left( 4\,a\,B_{{4}}\,\sin \,2\theta  +4\, a\,B_{{3}}\,\cos\,2\theta
-c_{{32}}\,\cos \,\theta + c_{{31}}\,\sin \,\theta \right) \,T''
\nonumber \\&
-2\, \left( 2\,a\,B_{{3}}\,\sin
 \,2\theta  -2\,a\,B_{{4}}\,\cos \,2\theta  -\,c_{{31}}\,\cos \,\theta
- c_{{32}}\,\sin \,\theta \right) \, T^{(3)}
\nonumber \\&
-6\,a \left( c_{{12}}\,\sin \,2\theta +c_{{11}}\,\cos \,2\theta   \right) \ .
\label{EqVr}
\end{align}

For $a=0$ (\ref{EqVr}) coincides with (\ref{Ctb}).

\textbf{Case I}. $B_{3}=B_{4}=c_{11}=c_{12}=0$, $A_{1}$ and $A_{2}$ arbitrary.
The non-linear equation
(\ref{Ctc}) is satisfied trivially, while (\ref{EqVr}) coincides with
(\ref{Tnle5theta}) and thus, in this case, we obtain the quantum potentials

\begin{equation}
\begin{aligned}
 V(r,\theta)& = \ \frac{a}{r}\ + \ \frac{\partial_{\theta}T(x_{\pm})}{r^2} \\ &
= \ \frac{a}{r}\ + \ \frac{\hbar^2}{r^2}\Bigg(\,W'(x_{\pm}) \mp  \frac{2\,\cos
\theta}{\sin^2\theta }  \,W(x_{\pm})
+ \frac{1}{2\,\sin^2\theta }\,\Gamma  \Bigg) \ ,
\end{aligned}
\end{equation}

where $\Gamma=(\gamma_2+\gamma_4+\sqrt{2\,\gamma_1}-\gamma_1-\gamma_3-\frac{3}{8})$,
and both $T$ and $W$ are completely defined through (\ref{transf})-(\ref{ParcA}).
These potentials correspond to the integral $(A_{2}=1)$
\begin{equation}
 \begin{split}
 Y\ =& \    \hbar^4\,\{ \partial_\theta^3,\, \sin\theta\, \partial_r \}\,
+  \frac{\hbar^4}{r}\{ \partial_\theta^3,\,  \cos\theta \partial_\theta\}  -
 \ \hbar^2\,\{ G_1(r,\theta),\,\partial_r^2  \} \ - \ \hbar^2\, \{ G_3(r,\theta),\,\partial_r\,\partial_\theta  \}
\\&
\ - \  \hbar^2\,\{ G_2(r,\theta),\,\partial_\theta^2  \}
\ + \  G_4(r,\theta)\  ,
\end{split}
\end{equation}
where
\begin{align}
 &   G_{{1}} \left( r,\theta \right) =0  \,, \\
 &   G_{{2}} \left( r,\theta \right)  =  \frac{1}{r} \,
\Big(
4\,\cos  \theta \ T'
 +2\,c_{{32}}\,\cos  \theta
- \left( T + 2\,c_{{31}}   \right) \sin  \theta
\Big) + a\,\cos \theta   \,, \nonumber \\
&
 G_{{3}} \left( r,\theta \right) = 3\,\sin  \theta \ T'   +
 \left( T   +2\,c_{{31}} \right) \cos  \theta
  +2\,c_{{32}}\,\sin \theta \ \,,\nonumber \\
&
G_{{4}} \left( r,\theta \right) =  \,
{\frac {1}{2\,r}} \Big( \,\sin  \theta \ T^{(4)}
+ 4\,\cos \theta  \  T^{(3)}
- 3\,\sin  \theta  \ T''
- 2\,\cos  \theta  \ T'
\Big) {{\hbar}}^{2}
\, \nonumber \\ &
-
{\frac {2}{{r}}} \Big(
  3\,\sin  \theta  \ T'
+ \,\cos  \theta  \ T
+ 2\,c_{{31}}  \cos  \theta
+ 2\,c_{{32}}  \sin  \theta     \Big) \,T''
\nonumber  \\ &
  - 2\,a\,\sin  \theta \ T
 +4\, a\,\cos  \theta \ T'  -4\,a\,(c_{31}\,\sin\theta-c_{32}\,\cos\theta) - a\,{{\hbar}}^{2}\,\cos \theta\,  \ ,
\nonumber
\end{align}
here $T'=\partial_{\theta}T(x_{\pm})$.

\textbf{Case II}. For $A_{1}=A_{2}=c_{31}=c_{32}=0$, $a\neq0$ and $B_{3},\,B_4$
arbitrary
(\ref{EqVr}) reduces to a linear equation. For exotic potentials it must be
satisfied identically. This implies $B_3=B_4=0$, so no fourth order integral
exists.

\section{CLASSICAL POTENTIALS}
\label{CLASSICAL POTENTIALS}

The two determining equations (\ref{Ctb})-(\ref{Ctc}) reduce to third order
equations for $T(\theta)$ once we impose the condition $\hbar\rightarrow 0$. The
limit is singular and interestingly, the equations in this case do not pass the
Painlev\'e test. The division into subcases (\ref{Cases}) remains. We can always
integrate (\ref{Ctb})-(\ref{Ctc}) twice and we obtain a first order nonlinear
equation of the form

\begin{equation}
\label{ME}
Q_4(z)\,{T'}^2 + Q_1(z)\,T\,T' + T^2 +Q_2(z)\,T' + c\,T + Z(z) \ = \ 0 \ ,
\end{equation}
where $Q_n(z)$ is a polynomial in $z$ of order $n$, $Z(z)$ is a rational
function and $c$ is a constant. Using the transformation
\[
T(z) \ = \ m(z)\,t(z) + n(z) \ ,
\]
we can factorize (\ref{ME}) as follows
\[
(t'-t'_0)(t'+t'_0)  \ =\ 0 \ ,
\]
where
\[
t'_0 \ = \ \frac{\sqrt{\left(Q_1\, (m \,t+n) + Q_2\right){}^2-4\, Q_4 \left(Q_0
(m\, t+n)+(m \,t+n)^2+Z\right)}}{2\, m\, Q_4} \ .
\]
and $m$ and $n$ satisfy
\[
2 \,Q_4(z) \,n'(z) + n(z)\, Q_1(z) + Q_2(z) \ = \ 0 \ \ ,
\]
\[
 2\,Q_4(z)\,m'(z) + m(z)\,Q_1(z)  \ =\ 0 \ ,
\]

In general, explicit solutions to the equation $t'\pm t'_0=0$ are not known.
However for special values of the parameters contained in the $Q_i$ and $Z$, the
function $t'_0$ becomes linear in $t$ and explicit solutions can be constructed.

\subsection{Case $V(r,\,\theta)= \frac{S(\theta)}{r^2}$}

\subsubsection{Case I}

The classical potential $S(\theta)=T'(\theta)$ satisfies (\ref{NLE1}) with
$\hbar\mapsto0$. This limit   is singular, the order
of the
equation (\ref{NLE1}) drops from three to one.  The so obtained
non-linear
first order differential equation reads:
\begin{align}
& T ^{2}- 2\left( \,z\, T' - 2\,{c}_{31}\right) T
-3\,{z}^{2} \left( {z}^{2}+1 \right) \left( T'\right)^{2}
- 4\,z \left( {c}_{32}\,z + {c}_{31} \right) \,T'
\nonumber \\&
 +2\,{\frac {{{K}_1}}{\sqrt {{z}^{2}+1}}}   -  \frac{K_2}{2} \ =\ 0\ ,
\label{class1}
\end{align}

\noindent where $z=\tan\ \theta$. Factorization of the l.h.s in (\ref{class1}) in the form
of a
product of
two factors of first order allows us to find particular solutions.
These two factors become linear for specific values of the parameters in
(\ref{class1}) only.  Namely,  putting ${K}_1={c}_{32}=0$ and $K_2=-8\,
{{c}_{31}}^{2}$ in (\ref{class1}) we obtain the equation

\[
6\,z^2(1+z^2)\,\bigg(T'+\frac{(2\,c_{31}+T)(1+\sqrt{4+3\,z^2})}{3\,z\,(1+z^2)}
\bigg)
\bigg(T'+\frac{(2\,c_{31}+T)(1-\sqrt{4+3\,z^2})}{3\,z\,(1+z^2)}\bigg) = 0 \ ,
\]

from which we derive two particular solutions:

\begin{align}
T_1&=  -2\,{c}_{31} + \alpha
\,\frac{z^{\frac{1}{3}}\,{(3 z^2+2 \sqrt{3 z^2+4}+5)}^{\frac{1}{6}}}{ {(\sqrt{3
z^2+4}+2)}^{\frac{2}{3}}}\,, \nonumber   \\[10pt]
T_2&=  -2\,{c}_{31} +  \alpha
\,\frac{{(1+z^2)}^{\frac{1}{3}}\,{(2+\sqrt{4+3\,z^2})}^{\frac{2}{3}}}{z\,{(3
z^2+2 \sqrt{3 z^2+4}+5)}^{\frac{1}{6}}} \ ,
\label{Tcas1}
\end{align}
where $\alpha$ is an integration constant.
By differentiating the preceding results (\ref{Tcas1}) with respect to $\theta$
we obtain the classical potentials:
{\scriptsize
\begin{equation}
 V_1(r,\,\theta)=\frac{3\alpha\sec^4\theta\Big[7+3\sqrt{4+3\tan^2\theta}
+\cos2\theta\big(1+\sqrt{4+3\tan^2\theta}\big)\Big]}
{r^2\tan^{\frac{2}{3}}\theta\sqrt{4+3\tan^2\theta}\Big(2+\sqrt{4+3\tan^2\theta}
\Big)^{\frac{5}{3}}\Big(5+3\tan^2\theta+2\sqrt{4+3\tan^2\theta}\Big)^{\frac{5}{6
}}}
\label{V1clas} \ ,
\end{equation}
}
and
{\scriptsize
\begin{equation}
 V_2(r,\,\theta)=-\frac{\alpha\sec^{\frac{2}{3}}\theta\Big[47+17\sqrt{
4+3\tan^2\theta}+18\cot^2\theta(2+\sqrt{4+3\tan^2\theta})+3\tan^2\theta(5+\sqrt{
4+3\tan^2\theta})\Big]}
{2r^2\sqrt{4+3\tan^2\theta}\Big(2+\sqrt{4+3\tan^2\theta}\Big)^{\frac{1}{3}}
\Big(5+3\tan^2\theta+2\sqrt{4+3\tan^2\theta}\Big)^{\frac{7}{6}}}
\label{V2clas}\ .\\
\end{equation}
}

In general, the potentials $V(r,\,\theta)$ associated with ({\ref{class1}})
possess the integral
\begin{equation}
 \begin{split}
 Y =&  2 \, \sin\theta \,p_\theta^3\,p_r
+  \frac{2}{r}\,\cos\theta \,p_\theta^4  +
 \ 2\, G_1(r,\theta)\,p_r^2   \ + \  2\, G_3(r,\theta)\,p_r\,p_\theta
\\&
\ + \  2\, G_2(r,\theta)\,p_\theta^2
\ + \  G_4(r,\theta)\  ,
\label{Ycase12class}
\end{split}
\end{equation}
 ($A_2=1$) where
\begin{align}
 &   G_{{1}} \left( r,\theta \right) =0  \,, \\
 &   G_{{2}} \left( r,\theta \right)  =  \frac{1}{r} \,
\Big(
4\,\cos  \theta \ T'
 +2\,c_{{32}}\,\cos  \theta
- \left( T + 2\,c_{{31}}   \right) \sin  \theta
\Big)   \,, \nonumber \\
&
 G_{{3}} \left( r,\theta \right) = 3\,\sin  \theta \ T'   +
 \left( T   +2\,c_{{31}} \right) \cos  \theta
  +2\,c_{{32}}\,\sin \theta  \ ,\nonumber \\
&
G_{{4}} \left( r,\theta \right) =  -{\frac {2}{{r}}} \Big(
  3\,\sin  \theta  \ T'
+ \,\cos  \theta  \ T
+ 2\,c_{{31}}  \cos  \theta
+ 2\,c_{{32}}  \sin  \theta     \Big) \ T''
 \ .\nonumber
\end{align}

\subsubsection{Case II}

\bigskip

The classical potential $S(\theta)$ satisfies (\ref{NLE2}) with
$\hbar\mapsto0$. This limit   is singular, the order
of the
equation (\ref{NLE2}) drops from three to one.  The so obtained
non-linear
first order differential equation in $T$ reads:
\begin{align}
&   T^{2}  - \left(  {c}_{11} + 2\,z\, T'   \right) T
 -3\,{z}^{2} \left( {z}^{2}+1 \right)  \left( T' \right)^{2}
+   z \left( {c}_{12} \,z + {c}_{11} \right) \, T'
\nonumber \\&
- \frac{1}{4}\,\frac {{K_1}}{\sqrt {{z}^{2}+1}}    +{K_2} =0 \,,
\label{class2}
\end{align}

Factorization of the l.h.s in (\ref{class2}) in the form of a
product of
two factors of first order allows us to find particular solutions again.
The factors are linear for specific values of the parameters in
(\ref{class2}) only. These special values  are ${K}_1={c}_{12}=0$ and
$K_2=\frac{1}{4} {{c}_{11}}^{\,2}$.
By substituting these values in (\ref{class2}) we derive two particular
solutions
\begin{align}
T_3&=  \frac{{c}_{11}}{2} + \alpha
\,\frac{z^{\frac{1}{3}}\,{(3 z^2+2 \sqrt{3 z^2+4}+5)}^{\frac{1}{6}}}{ {(\sqrt{3
z^2+4}+2)}^{\frac{2}{3}}}\,, \nonumber   \\[10pt]
T_4&=  \frac{{c}_{11}}{2} +  \alpha
\,\frac{{(1+z^2)}^{\frac{1}{3}}\,{(2+\sqrt{4+3\,z^2})}^{\frac{2}{3}}}{z\,{(3
z^2+2 \sqrt{3 z^2+4}+5)}^{\frac{1}{6}}} \ ,
\label{Tcas2}
\end{align}
where $\alpha$ is an integration constant. By differentiating the preceding
results (\ref{Tcas2}) with respect to $\theta$ we obtain the classical
potentials:

{\scriptsize
\begin{equation}
\label{V3clas}
V_{3}(r,\,\theta)=\frac{3\alpha\sec^42\theta\Big[7+3\sqrt{4+3\tan^22\theta}
+\cos4\theta\big(1+\sqrt{4+3\tan^22\theta}\big)\Big]}
{ r^2\,
\tan^{\frac{2}{3}}2\theta\sqrt{4+3\tan^22\theta}\Big(2+\sqrt{4+3\tan^22\theta}
\Big)^{\frac{5}{3}}\Big(5+3\tan^22\theta+2\sqrt{4+3\tan^22\theta}\Big)^{\frac{5}
{6}}}
 \, ,
\end{equation}
}
and
{\scriptsize
\begin{equation}
\label{V4clas}
V_4(r,\,\theta)=-\frac{\alpha\sec^{\frac{2}{3}}2\theta\Big[47+17\sqrt{
4+3\tan^22\theta}+18\cot^22\theta(2+\sqrt{4+3\tan^22\theta}
)+3\tan^22\theta(5+\sqrt{4+3\tan^22\theta})\Big]}
{2\,r^2\,\sqrt{4+3\tan^22\theta}\Big(2+\sqrt{4+3\tan^22\theta}\Big)^{\frac{1}{3}
}\Big(5+3\tan^22\theta+2\sqrt{4+3\tan^22\theta}\Big)^{\frac{7}{6}}}\,,
\end{equation}
}

where $\alpha$ is a constant.

For ({\ref{class2}}) the potentials $V(r,\,\theta)$ possess the integral

\vspace{0.2cm}

{\footnotesize
\begin{equation}
\begin{split}
 Y  =&
  2\, p_r^2 \,  p_\theta^2\,   \cos 2\theta  -  \frac{2}{r^2} \,  p_\theta^4\,
\cos 2\theta   - \frac{4}{r}\,p_r \, p_\theta^3\, \sin 2\theta
\\&
+
\ 2\, G_1(r,\theta)\,p_r^2   \ + \  2\, G_3(r,\theta)\,p_r\,p_\theta
\ + \  2\, G_2(r,\theta)\,p_\theta^2
\ + \  G_4(r,\theta)\  ,  \label{Ycase34clas}
\end{split}
\end{equation}
}

(with $ B_{3}=1$) where
\begin{align*}
&
  G_{{1}} \left( r,\theta \right) =
2\,\cos  2\,\theta \  T'
- 2\,\sin  2\,\theta \ T
+  c_{{11}} \,\sin  2\,\theta
- c_{{12}} \,\cos  2\,\theta
\nonumber \,, \\ &
G_{{2}} \left( r,\theta \right) = {\frac {1}{{r}^{2}}} \Big(
  2\,\sin  2\,\theta \ T
- 4\,\cos  2\,\theta \ T'
-
  c_{{11}} \,\sin  2\,\theta
+ c_{{12}} \,\cos  2\,\theta
   \Big)  \,,
\nonumber  \\ &
  G_{{3}} \left( r,\theta \right) = \frac{1}{r}
  \Big(    2\,c_{{11}}\,\cos  2\,\theta     - 4\,\cos  2\,\theta \  T
  - 6\,\sin  2\,\theta \  T'
+  2\,c_{{12}}\,\sin  2\,\theta
\Big)
\nonumber \,, \\ &
  G_{{4}} \left( r,\theta \right) =
\frac {1}{{r}^{2}}\, \Big(
  \, \left(
    \left( 4\,T   - 2\,c_{{11}} \right) \cos  2\,\theta
 + \left( 6\,T' -  2\,c_{{12}} \right) \sin  2\,\theta
     \right) \,T''
 + \,\left( 8\,T'   - 4\,c_{{12}} \right) \, T'\,   \cos  2\,\theta
 \nonumber \\&
 - \, \left( 8\,T   - 4\,c_{{11}} \right) \, T'\, \sin  2\,\theta
  \Big)
   \ .
\end{align*}

\bigskip

\subsection{Potential $V(r,\,\theta) = b\,r^2 + \frac{S(\theta)}{r^2}$}
The classical potentials are given by
\begin{equation}\label{Vparah}
V(r,\,\theta)\ = \   b\,r^2 + \frac{T'(\theta)}{r^2} \ ,
\end{equation}
$T$ from (\ref{class2}), and they correspond to the integral
{\footnotesize
\begin{align}
 Y \ = \  &
 2\, p_r^2 \,  p_\theta^2\,   \cos 2\theta  -  \frac{2}{r^2} \,  p_\theta^4\,
\cos 2\theta   - \frac{4}{r}\,p_r \, p_\theta^3\, \sin 2\theta
\\&
+
\ 2\,G_1(r,\theta)\,p_r^2  \ + \  2\, G_3(r,\theta)\,p_r\,p_\theta
\ + \  2\, G_2(r,\theta)\,p_\theta^2
\ + \  G_4(r,\theta)\ .
\end{align}
}

\vspace{0.2cm}

(with $ B_{3}=1$) where

\begin{align}
&
  G_{{1}} \left( r,\theta \right) =
2\,\cos  2\,\theta \ T'
- 2\,\sin  2\,\theta \ T
+  c_{{11}} \,\sin  2\,\theta
- c_{{12}} \,\cos  2\,\theta
\nonumber \,, \\ &
G_{{2}} \left( r,\theta \right) = {\frac {1}{{r}^{2}}} \Big(
  2\,\sin  2\,\theta \ T
- 4\,\cos  2\,\theta \ T'
-
  c_{{11}} \,\sin  2\,\theta
+ c_{{12}} \,\cos  2\,\theta
  \Big) + 2\,b\,r^2\,\cos 2\,\theta    \,,
\nonumber  \\ &
  G_{{3}} \left( r,\theta \right) = \frac{1}{r}
  \Big(    2\,c_{{11}}\,\cos  2\,\theta     - 4\,\cos  2\,\theta \  T
  - 6\,\sin  2\,\theta \  T'
+  2\,c_{{12}}\,\sin  2\,\theta
\Big)
\nonumber \,, \\ &
  G_{{4}} \left( r,\theta \right) =
\frac {1}{{r}^{2}}\, \Big(
  \, \left(
    \left( 4\,T   - 2\,c_{{11}} \right) \cos  2\,\theta
 + \left( 6\,T' -  2\,c_{{12}} \right) \sin  2\,\theta
     \right) \ T''
 + \,\left( 8\, T'   - 4\,c_{{12}} \right) \ T'   \cos  2\,\theta
 \nonumber \\&
 - \, \left( 8\,T   - 4\,c_{{11}} \right) \ T' \sin  2\,\theta
  \Big) + 4\,b\,r^2\,(2\,\cos  2\,\theta\,T'-2\,\sin  2\,\theta\,T+c_{11}\,\sin
2\,\theta-c_{12}\,\cos  2\,\theta)
  \ .
\end{align}

\subsection{Potential $V(r,\,\theta) = \frac{a}{r} + \frac{S(\theta)}{r^2}$}

Similarly, the classical potentials are given by

\begin{equation}
V(r,\,\theta) = \frac{a}{r} + \frac{T'(\theta)}{r^2} \ ,
\end{equation}

$T$ from (\ref{class1}), and they corresponds to the integral

\begin{equation}
 \begin{split}
 Y \ = \ &   2 \, \sin\theta \,p_\theta^3\,p_r
+  \frac{2}{r}\,\cos\theta \,p_\theta^4  +
 \ 2\, G_1(r,\theta)\,p_r^2   \ + \  2\, G_3(r,\theta)\,p_r\,p_\theta  +
\\&
\ + \ 2\, G_2(r,\theta)\,p_\theta^2
\ + \  G_4(r,\theta)\  ,
\end{split}
\end{equation}

 ($A_2=1$) where
\begin{align}
 &   G_{{1}} \left( r,\theta \right) =0  \,, \\
 &   G_{{2}} \left( r,\theta \right)  =  \frac{1}{r} \,
\Big(
4\,\cos  \theta \, T'
 +2\,\cos  \theta \,c_{{32}}
- \left( T + 2\,c_{{31}}   \right) \sin  \theta
\Big) +a\,\cos \theta \,, \nonumber \\
&
 G_{{3}} \left( r,\theta \right) = 3\,\sin  \theta \,T'   +
 \left( T   +2\,c_{{31}} \right) \cos  \theta
  +2\,\sin \theta \,c_{{32}} \,,\nonumber \\
&
G_{{4}} \left( r,\theta \right) =  -{\frac {2}{{r}}} \Big(
  3\,\sin  \theta  \,T'
+ \,\cos  \theta  \,T
+ 2\,c_{{31}}  \cos  \theta
+ 2\,c_{{32}}  \sin  \theta     \Big) \,T''
\nonumber
\\ &  + 2\,a\,(2\,\cos \theta \,T'-\sin \theta
\,T)-4\,a\,(c_{31}\,\sin\theta-c_{32}\,\cos\theta)
 \ . \nonumber
\end{align}

\clearpage

\section{POLYNOMIAL ALGEBRA}
\label{POLYNOMIAL ALGEBRA}

In this section we discuss the algebra of the integrals of
motion in the classical case~\cite{Daskaloyannis:2006,Daskaloyannis:2007,Hoque2015}.

Take the second order integral $X$ and the fourth order ones $Y$, (\ref{X})
and (\ref{Y}) respectively. Let us define, via their Poisson bracket $\{\}_{{}_{PB}}$, the fifth
order polynomial in momenta
\begin{align}
C \ \equiv \{ \,Y,\,X\,\}_{{}_{PB}} \ ,
\end{align}

which by construction is also an integral of motion. Now we study the algebra generated by the four quantities $H,\,X,\,Y\,$ and $C$. The relevant (non vanishing) Poisson brackets are $\{ \,X,\,C\, \}_{{}_{PB}}$ and $\{ \,Y,\,C\, \}_{{}_{PB}}$ only.

First we consider the case of the extended harmonic oscillator potential

\[ V(r\,,\theta) \ =\
b\,r^2 + \frac{T'(\theta)}{r^2}\ . \]

For the particular solutions (\ref{Tcas2}), $T_3$ and $T_4$, the algebra generated by the integrals is given by
\begin{align}
\{ \,X,\,C\, \}_{{}_{PB}} \ &=\ 16\,X\,Y   \ , \\
\nonumber
\{ \,Y,\,C\, \}_{{}_{PB}}\ &=\ 8\,\bigg[
48\,{H}^{2}{X}^{2}-128\,{X}^{3}\,b -{Y}^2 + b\,\sigma_{3,4} \ \bigg]
\ ,
\end{align}
where $\sigma_3=\frac{512}{9}\, \alpha^3$, $\sigma_4=-\frac{512}{3}\,
\alpha^3$ and $\alpha$ a non zero constant, respectively. At $b=0$ this algebra reduces to that of the Case II, $R(r)=0$. For an arbitrary solution of (\ref{class2}), in order to the algebra to be closed the function $T$ must satisfy a sixth order polynomial equation presented in the Appendix B. Then the algebra takes the form

\begin{align}
\{ \,X,\,C\, \}_{{}_{PB}} \ &=\ 16\,X\,Y  -32\,K_1\,H \ , \\
\nonumber
\{ \,Y,\,C\, \}_{{}_{PB}}\ &=\ 8\,\bigg[
48\,{H}^{2}{X}^{2} -{Y}^2 -128\,b\,{X}^{3} -64\,c_{12}\,H^2\,X   +16(c_{11}^2+c_{12}^2-4\,K_2)H^2 +192\,b\,c_{12}\,X^2  \ \bigg]
\\
\nonumber
& \qquad   - 512\,b\,(c_{11}^2+c_{12}^2-4\,K_2)\,X - b\,\lambda  \ ,
\end{align}

where $\lambda$ is an arbitrary constant. It is a quartic polynomial algebra.

For the extended Coulomb potential \[ V(r\,,\theta) \ =\  \frac{a}{r} +
\frac{T'(\theta)}{r^2}\ , \] with the particular solutions $T_1$ and $T_2$ we
have that
\begin{align}
\{ \,X,\,C\, \}_{{}_{PB}} \ &=\ 4\,X\,Y   \ , \\
\nonumber
\{ \,Y,\,C\, \}_{{}_{PB}}\ &=\ 2\,\bigg[ 32\,H\,X^3 + 12\,a^2\,X^2 - Y^2 + \sigma_{1,2}\,H
\ \bigg]
\ ,
\end{align}
where $\sigma_1=-\frac{16}{9}\, \alpha^3$ and $\sigma_2=\frac{16}{3}\,
\alpha^3$, respectively. At $a=0$ this algebra corresponds to the Case I, $R(r)=0$. Similarly, for a general solution of (\ref{class1}) the function $T$ must also satisfy a sixth order polynomial equation and the corresponding algebra reads

\begin{align}
\{ \,X,\,C\, \}_{{}_{PB}} \ &=\ 4\,X\,Y  + 8\,a\,K_1 \ , \\
\nonumber
\{ \,Y,\,C\, \}_{{}_{PB}}\ &=\ 2\,\bigg[
32\,{H}{X}^{3} -{Y}^2 + 12\,a^2\,{X}^{2} + 96 \,c_{32}\,H\,X^2 + 64(c_{31}^2+c_{32}^2+\frac{K_2}{8})H\,X + 32\,a^2\,c_{32}\,X  \ \bigg]
\\
\nonumber
& \qquad   - \lambda\,H + 32\,a^2\,(c_{31}^2+c_{32}^2+\frac{K_2}{8})   \ ,
\end{align}

In the classical case the algebra of $H,\,X,\,Y\,$ and $C$ is useful to obtain and classify the
trajectories. In full generality, namely for general solutions of (\ref{class1})
and (\ref{class2}), an algebraic equation for the non-trivial part $T(\theta)$
of the potential can be derived by requiring the algebra to be closed.

In the quantum case, once the functions $G_1,...,G_4$ (\ref{Gfunctionsg}) figuring in the integral $Y$ (\ref{YA}) are calculated, it is possible to express the two commutators $[ \,X,\,C\, ]$ and $[ \,Y,\,C\, ]$ as polynomials in $X,Y$ and $H$. As a matter of fact, the condition that the algebra of the integrals of motion should close leads directly to the fifth order equations (\ref{Tnle5theta}) and (\ref{Tnle52theta}) for $T$. Moreover, this closure also provides the integrals of these equations such as e.g. eq. (\ref{Sec}).

\section{CONCLUSIONS}
\label{Conclusions}

We studied superintegrability in a two-dimensional Euclidean space. Classical
and quantum fourth-order superintegrable potentials separating in polar
coordinates were derived. We can summarize the main results via the following
Theorems

{\bf Theorem 1.} \emph{In quantum mechanics, the confining superintegrable systems
correspond to }
\[
V(r,\,\theta) \ = \ \frac{a}{r} + \frac{\hbar^2}{r^2}\Bigg(\,W'(x_{\pm}) \mp
\frac{2\,\cos \theta}{\sin^2\theta }  \,W(x_{\pm})
+
\frac{8(\gamma_2+\gamma_4+\sqrt{2\,\gamma_1}-\gamma_1-\gamma_3)-3}{16\,
\sin^2\theta } \Bigg) \ ,
\]
\emph{here} $x_{\pm}=\sin^2(\frac{\theta}{2}),\cos^2(\frac{\theta}{2})$
\emph{and}
\[
V(r,\,\theta) \ = \ b\,r^2 +  \frac{\hbar^2}{r^2}\Bigg(\,4\,W'(x_{\pm}) \mp
\frac{8\,\cos 2\theta}{\sin^2 2\theta }  \,W(x_{\pm})
+ \frac{4(\gamma_2+\gamma_4+\sqrt{2\,\gamma_1}-\gamma_1+\gamma_3)-3}{2\,\sin^2
2\theta }\Bigg) \ ,
\]
$x_{\pm}=\sin^2\theta,\cos^2\theta$ \emph{where} $W(x)$ \emph{is given by}
(\ref{Wpot}) \emph{in both cases.} \emph{The leading term of the integral Y in}
(\ref{YA}) \emph{is} $\{L_z^3,\,p_y\}$ \emph{and} $\{L_z^2,\,p^2_x-p_y^2\}$,
\emph{respectively.}

The non-confining potentials are given by (\ref{VcaseI}) with integral (\ref{YcaseI}), and (\ref{VWA7}) with integral (\ref{YWA7}).

The function
\[
W\ = \ W(x,\,P_6(x);\,\gamma_1,\,\gamma_2,\,\gamma_3,\,\gamma_4)
\]
is expressed in terms of the sixth Painlev\'e transcendent $P_6$ (\ref{P6}) in
full generality. In the case of a third order superintegrable system, not all
four $(\gamma_1,\,\gamma_2,\,\gamma_3,\,\gamma_4)$ but three constants in
(\ref{P6}) are arbitrary only. Moreover, the third order system does not allow
any confining potentials.

\bigskip

{\bf Theorem 2.} \emph{In classical mechanics, the superintegrable confining systems
correspond to }
\[
V(r,\,\theta) \ = \ \frac{a}{r} + \frac{T'(\theta)}{r^2} \ ,
\]
\emph{where} $T'$ \emph{satisfies} (\ref{class1}) \emph{and $a$ is an arbitrary
constant. The leading term of the integral Y in} (\ref{YA}) \emph{is}
$\{L_z^3,\,p_y\}$, \emph{and}
\[
V(r,\,\theta) \ = \ b\,r^2 + \frac{T'(\theta)}{r^2} \ ,
\]
\emph{here} $T'$ \emph{satisfies (\ref{class2}), $b$ is constant, and the
leading term of Y is given by } $\{L_z^3,\,p^2_x-p_y^2\}$.

Particular solutions of (\ref{class1}) and (\ref{class2}) were presented in
(\ref{Tcas1}) and (\ref{Tcas2}), respectively\,.

The non-confining superintegrable systems are given by (\ref{V1clas}) and (\ref{V2clas}) with integral (\ref{Ycase12class}), and (\ref{V3clas}), (\ref{V4clas}) with integral (\ref{Ycase34clas}), respectively.

\bigskip

Work is currently in progress on a continuation of this article. We will
add a general investigation of the polynomial algebra generated by the
integrals of motion in the classical and quantum cases.
We also plan to present figures of the classical trajectories and to use
the algebra of integrals to calculate the energy spectrum and the wave
functions in the quantum case. Another part of the project is to determine
all  corresponding non-exotic potentials.

\section{ACKNOWLEDGMENTS}

The research of P. W. was partially  supported by a research grant from NSERC of
Canada. J.C.L.V. thanks PASPA grant (UNAM, Mexico) and the Centre de Recherches
Math\'ematiques, Universit\'e de Montr\'eal for the kind hospitality while on
sabbatical leave during which this work was done.  The research of A.M.E. was
partially supported by a fellowship awarded by the Laboratory of Mathematical Physics of the CRM and by
CONACyT grant 250881 (Mexico) for postdoctoral research.

\appendix

\section{Functions $F_i$}

Explicitly, the functions $F_{1}, \ldots ,F_{13}$ in (\ref{Eq30})-(\ref{Eq01}) are given by
\begin{align}
& F_1 = 2 \Big(
\,  { B}_{{1}}\,\cos 2\,\theta
+\, { B}_{{2}}\,\sin  2\,\theta
+\, { D}_{{1}}\,\cos  4\,\theta
+\, { D}_{{2}}\,\sin  4\,\theta
\Big)\,,
 \\[10pt]  &
F_2 =
  \frac{1}{r}
\Big({B}_{{2}}\,\cos 2\,\theta
-    {B}_{{1}}\,\sin 2\,\theta
- 2\,{D}_{{1}}\,\sin 4\,\theta
+ 2\,{D}_{{2}}\,\cos 4\,\theta
\Big)  \nonumber \\ &
+     {A}_{{3}}\,\cos\, \theta
+              {A}_{{4}}\,\sin\, \theta
+   {C}_{1}\,  \sin 3\,\theta
+              {C}_{2}\,  \cos 3\,\theta
\,,
\nonumber
\end{align}
%
\begin{align}
F_3 &=
 \frac{1}{ {r}^{3} } \Big(
   {B}_{{2}}\,\cos  2\theta
-  {B}_{{1}}\,\sin  2\theta
+ 2{D}_{{1}}\,\sin  4\theta
- 2{D}_{{2}}\,\cos  4\theta
\Big)
\nonumber
\\ &
+ \frac{1}{{r}^{2}}
\Big(
  {A}_{{4}}\,\sin \theta
+ {A}_{{3}}\,\cos \theta
- 3\, {C}_{2} \cos  3\,\theta
- 3\, {C}_{1} \sin  3\,\theta
\Big)
\nonumber
\\ &
- \frac{2}{{r}}
\Big(
{B}_{{3}} \,\sin 2\,\theta
-{B}_{{4}}\,\cos 2\,\theta
\Big)
+ {A}_{{2}}\,\sin \theta + {A}_{{1}}\,\cos \theta
\,,
\nonumber
\end{align}
%
\begin{align}
F_4 &=
\frac{2}{{r}^{4}}
\Big(
  \, {D}_{{1}}\,\cos 4\,\theta
+ \, {D}_{{2}}\,\sin 4\,\theta
- \, {B}_{{1}}\,\cos 2\,\theta
- \, {B}_{{2}}\,\sin 2\,\theta
\Big)
\nonumber \\ &
+ \frac{4}{{r}^{3}}
\Big(
  {A}_{{4}}\,\cos \theta
- {A}_{{3}}\,\sin \theta
- {C}_{1}\, \cos 3\,\theta
+ {C}_{2}\, \sin 3\,\theta
\Big)
\nonumber \\ &
- \frac{4}{{r}^{2}}
\Big(
  {B}_{{3}}\,  \cos  2\,\theta
+ {B}_{{4}} \,\sin  2\,\theta
\Big)
+ \frac{4}{{r}}
\Big(
  {A}_{{2}}  \,\cos \theta
- {A}_{{1}} \,\sin \theta
\Big)\,,
\nonumber
\end{align}
%
\begin{align}
F_5 &=
- \frac{6}{{r}^{2}}
\Big(
   \,{ D}_{{1}}\,\cos  4\,\theta
+  \,{ D}_{{2}}\,\sin  4\,\theta
\Big)
+ \frac{2}{{r}}
\Big(
    {A}_{{4}} \,\cos \theta
-   {A}_{{3}} \,\sin \theta
\nonumber \\&
+ 3(\,{C}_{1} \cos  3\,\theta
- {C}_{2} \sin  3\,\theta)
\Big)
+ 2\,({B}_{{3}} \,\cos  2\,\theta
+ {B}_{{4}} \,\sin  2\,\theta)
\,,
\nonumber
\end{align}
%
\begin{align*}
F_6 & \ = \
\frac{3}{r}
\left(
  B_2 \,{\cos 2\theta }
-B_1 \,{\sin 2\theta }
+2\,D_2 \,{\cos 4\theta }
-2 \,D_1 \,{\sin 4\theta }    \right)
\\ &
+3 \left(
   A_4 \,{\sin \theta }
+ A_3 \,{\cos \theta }
+ C_1 \,{\sin 3\theta }
+ C_2 \,{\cos 3\theta } \right)
\, ,
\nonumber
\end{align*}

\clearpage

\begin{align*}
F_7 &  \ =
\frac{12}{r^2}
\left(
  \,D_1 \,{\sin 4\theta }
- \,D_2 \,{\cos 4\theta }
\right)
 +  \frac{1}{2 \,r} \left(
  A_4 \,{\sin \theta }
+  A_3 \,{\cos \theta }
\right.\\& \left.
-30 \,C_1 \,{\sin 3\theta }
-30 \,C_2 \,{\cos 3\theta }  \right)
-  2\,(B_3 \,{\sin 2\theta}
-B_4 \,{\cos 2\theta})
\, ,
\nonumber
\end{align*}
\begin{align*}
F_8 &  \ = \  -\frac{3}{r^3} \left(
   \,B_1 \,{\cos 2\theta }
+  \,B_2 \,{\sin 2\theta }
-5 \,D_1 \,{\cos 4\theta }
-5 \,D_2 \,{\sin 4\theta }
\right)
\\ &  + \frac{3}{2 r^2} \left(
   A_4 \,{\cos \theta }
-  A_3 \,{\sin \theta }
- 9\,C_1 \,{\cos 3\theta }
+ 9\,C_2 \,{\sin 3\theta }
\right)
\\ &   -\frac{5}{r}  \left(
  B_3 \,{\cos 2\theta}
+ B_4 \,{\sin 2\theta}
\right)
+\frac{3}{2} \left(
   A_2 \, {\cos \theta }
-  A_1 \, {\sin \theta } \right)
\, ,
\nonumber
\end{align*}

\begin{align*}
F_9 &  \ = \  \frac{9}{r^4} \left(
   B_1 \,{\sin 2\theta }
-  B_2 \,{\cos 2\theta }
-2 \,D_1 \,{\sin 4\theta }
+2 \,D_2 \,{\cos 4\theta }
\right)
\\ &  +\frac{15}{2 r^3} \left(
3\,C_1 \, {\sin 3\theta }
+3\,C_2 \, {\cos 3\theta }
- A_3 \,{\cos \theta }
- A_4 \,{\sin \theta }
\right)
\\ &   +\frac{12}{r^2} \left(
   B_3 \,{\sin 2\theta }
-  B_4 \,{\cos 2\theta }
\right)
 -\frac{9}{2 \,r} \left(
 A_1 \, {\cos \theta }
 + A_2 \, {\sin \theta }
\right)
\,,
\nonumber
\end{align*}

\begin{align*}
F_{10} &  \ = \  -\frac{9}{r} \,\left(
  \,{D_1}\,{\cos 4\theta }
+ \,{D_2}\,{\sin 4\theta } \right)
+\frac{3}{2} \,(
   {A_4}\,{\cos \theta }
-  {A_3}\,{\sin \theta }
\\&
+ 3\, C_1\, {\cos 3\theta}
- 3 \,C_2\, {\sin 3\theta })
\,,
\nonumber
\end{align*}

\begin{align*}
F_{11} &  \ = \ -\frac{3}{r^2} \left(
    B_1\, {\cos 2\theta }
+   B_2 \,{\sin 2\theta }
- 5 \,D_1 \,{\cos 4\theta }
- 5\, D_2 \,{\sin 4\theta }
\right) - 4\, \left(
  B_3 \,  {\cos 2\theta }
+ B_4 \,  {\sin 2\theta }
\right)
\\ &  + \frac{1}{r} \left(
   A_4 \, {\cos \theta }
 - A_3 \, {\sin \theta }
- 9\,C_1 \,  {\cos 3\theta }
+ 9\,C_2 \,  {\sin 3\theta } \right)
\, ,
\nonumber
\end{align*}
\begin{align*}
F_{12} &  \ = \
\frac{2}{r^3}  \left(
  B_1 \,{\sin 2\theta }
- B_2 \,{\cos 2\theta }
- 14 \,D_1 \,{\sin 4\theta }
+ 14\,D_2 \,{\cos 4\theta }
\right)
\\ & +\frac{3}{2\, r^2}   \left(
 11\,C_1\, {\sin 3\theta }
+ 11\,C_2\, {\cos 3\theta }
-A_3 \,{\cos \theta }
- A_4 \,{\sin \theta }
\right)
\\ &   +\frac{6}{r}  \left(
  B_3 \,{\sin 2\theta }
- B_4 \,{\cos 2\theta }
\right) - \frac{3}{2} \left(
  A_1 \,{\cos \theta }
+ A_2 \,{\sin \theta }
 \right)
\, ,
\nonumber
\end{align*}
\begin{align*}
F_{13} &  \ = \  \frac{4}{r^4}   \left(
 2\, B_1\, {\cos 2\theta }
+2 \,B_2 \,{\sin 2\theta }
-11 \,D_1 \,{\cos 4\theta }
-11 \,D_2 \,{\sin 4\theta }
\right)
\\ & -\frac{2}{r^3}
\left(
  A_4  \,  {\cos \theta }
- A_3  \,  {\sin \theta }
- 17\,C_1 \,  {\cos 3\theta}
+ 17\,C_2 \,  {\sin 3\theta}
\right)
\\ &
+\frac{12}{r^2} \left(
   B_3\, {\cos 2\theta }
+  B_4\, {\sin 2\theta } \right)
- \frac{3}{r} \left(
   A_2 \, {\cos \theta }
-  A_1 \, {\sin \theta }
\right)
\ ,
\nonumber
\end{align*}

For a non-confining potential $R(r)=0$, the functions $F_i$ (\ref{Eq30p})-(\ref{Eq01p}) reduce to
\begin{align}
F_1&=0\,,  \nonumber \\
F_2&=0\,,  \nonumber \\
F_3&=A_{{1}} \cos \theta + A_{{2}} \sin \theta
+{\frac {2}{r}}
\big(
B_{4}  \cos 2\theta - B_{3} \sin 2\theta
\big) \,,
\nonumber \\
F_4&=
{\frac{4}{r}}
\big(
\,A_{{2}}  \cos \theta -\,A_{{1}} \sin \theta
\big)
-{\frac {4}{{r}^{2}}}
\big(
\,B_{{3}} \cos  2\theta  +\,B_{{4}}\sin  2\theta
\big)  \,,
\nonumber \\
 F_5&= 2(\,B_{{3}}\cos  2\theta  +\,B_{{4}}\sin  2\theta)
\,,
\nonumber \\
F_6 & \ = \  0 \,,
\nonumber \\
F_7 &  \ = 2\,(B_{4} \cos 2\theta \, - B_3\sin 2\theta) \, \,,
\nonumber \\
F_8 &  \ = \
\frac{3}{2} \left(A_2 \cos \theta  - A_{1} \sin \theta \right)
-\frac{5}{ r}   \left(B_3 \cos 2\theta + B_{4}\sin 2\theta   \right)\,,
\nonumber \\
F_9 &  \ = \
\frac{12}{r^2} \left( B_3\sin 2\theta  - B_{4} \cos 2\theta  \right)
 -\frac{9}{2\, r} \left(A_{1} \cos \theta + A_2\sin \theta \right) \,,
\nonumber \\
F_{10} &  \ = \  0
\,,
\nonumber \\
F_{11} &  \ = \ -4 \left(B_3 \cos 2\theta + B_{4}\sin 2\theta \right)\,,
\nonumber \\
F_{12} &  \ = \
\frac{6}{r} \left(B_3\sin 2\theta - B_{4}  \cos 2\theta \right)
- \frac{3}{2} \left( A_{1} \cos \theta + A_2 \sin \theta \right)
\,,
\nonumber \\
F_{13} &  \ = \
\frac{12}{r^2} \left(B_3 \cos 2\theta + B_{4}\sin 2\theta \right)
-\frac{3}{ r} \left(A_2 \cos \theta -A_{1} \sin \theta \right)
\,,
\nonumber
\end{align}

\section{Algebra of integrals of motion in classical limit}

An algebraic equation for the non-trivial part $T(\theta)$ of the potential was derived by requiring the algebra generated by the integrals of motion $H,\,X,\,Y\,$ and $C$ to be closed. For the extended harmonic oscillator potential

\[ V(r\,,\theta) \ =\
b\,r^2 + \frac{T'(\theta)}{r^2}\ , \]

the corresponding algebraic equation is a sixth order polynomial equation in $T$ given by

\begin{equation}
\tau_0(z) + \tau_1(z)\,T(z) + \tau_2(z)\,T^2(z)+ \tau_3(z)\,T^3(z)+ \tau_4(z)\,T^4(z)+ \tau_5(z)\,T^5(z)+ \tau_6(z)\,T^6(z) \ = \ 0 \  ,
\label{AleqA}
\end{equation}

where $z=\tan 2\,\theta$ and

\begin{equation*}
\begin{aligned}
\tau_0 \ = & \  786432 \, \left(c_{11}^4 \left(f K_1-4 K_2\right)+2 c_{11}^2 \left(K_1-4 f K_2\right){}^2+f K_1^3+48 f K_1 K_2^2-64 K_2^3-12 K_1^2 K_2\right)\,z^2
\\ & + 3072 \, c_{11} \left(-c_{11}^2 \left(256 c_{12} \left(16 K_2-3 f K_1\right)+\lambda \right)-\left(K_1-4 f K_2\right) \left(256 c_{12} \left(16 f K_2+5 K_1\right)+9 f \lambda \right)\right)\,z^3
\\ & + 3 \, \bigg(512 [-128 c_{12}^2 \left(-128 f^2 K_2^2-80 f K_2 K_1+K_1^2\right)
\\ & -6 c_{11}^2 \left(-256 c_{12}^2 \left(f K_1-8 K_2\right)+c_{12} \lambda -768 K_1 \left(K_1-4 f K_2\right)\right)
\\ &  +9 c_{12} f \lambda  \left(8 f K_2+K_1\right)-2048 K_2 \left(-12 f K_2 K_1+3 K_1^2+32 K_2^2\right)]+27 \lambda ^2\bigg)\,z^4
\\ & -1536 \, c_{11} \left(-512 c_{12}^3 \left(f K_1-16 K_2\right)+2304 c_{12} K_1 \left(8 f K_2+K_1\right)+6 c_{12}^2 \lambda +27 f \lambda  K_1\right)\,z^5
\\ &  3 \, \left(27 \left(65536 c_{11}^2 K_1^2+\lambda ^2\right)-1024 \left(c_{12}^3 \lambda +1024 c_{12}^4 K_2+16384 K_2^3\right)\right)\,z^6 \  ,
\end{aligned}
\end{equation*}

\begin{equation*}
\begin{aligned}
\tau_1 \ = & \   3145728 \, c_{11} \left(2 c_{11}^2 \left(f K_1-4 K_2\right)+c_{11}^4-8 f K_1 K_2+K_1^2+16 K_2^2\right)\,z^2
\\ &  -6144 \, [c_{11}^2 \left(256 c_{12} \left(11 f K_1+16 K_2\right)+15 \lambda \right) +f K_1 \left(1024 c_{12} K_2-9 \lambda \right)
\\ &  +4 K_2 \left(4096 c_{12} K_2+9 \lambda \right)-1280 c_{12} K_1^2-2048 c_{12} c_{11}^4]\,z^3  - 24576 \, c_{11} [-256 c_{12}^2 \left(3 c_{11}^2-4 f K_1+4 K_2\right)
\\ &  -128 \left(c_{11}^2 \left(9 f K_1-16 K_2\right)+12 f K_1 K_2-6 K_1^2+64 K_2^2\right)+3 c_{12} \lambda ]\,z^4
\\ &  + 3072 \, [4 c_{11}^2 \left(256 c_{12} \left(4 c_{12}^2+9 f K_1-32 K_2\right)-9 \lambda \right)-512 c_{12}^3 \left(f K_1-16 K_2\right)
\\ &  +256 c_{12} \left(72 f K_2 K_1+9 K_1^2-128 K_2^2\right)+6 c_{12}^2 \lambda +9 \lambda  \left(3 f K_1-8 K_2\right)]\,z^5
\\ &  -12288 \, c_{11} \left(9 c_{12} \lambda +4096 c_{12}^2 K_2-256 c_{12}^4+192 \left(9 K_1^2-64 K_2^2\right)\right)\,z^6     \  ,
\end{aligned}
\end{equation*}

\begin{equation*}
\begin{aligned}
\tau_2 \ = & \  -3145728 \, \left(2 c_{11}^2 \left(f K_1-4 K_2\right)+c_{11}^4-8 f K_1 K_2+K_1^2+16 K_2^2\right)\,z^2
\\ & + 294912 \,c_{11} \left(128 c_{12} \left(-c_{11}^2+f K_1+4 K_2\right)+\lambda \right)\,z^3
\\ & + 24576 \, [3 c_{12} \lambda -128 (c_{11}^2 \left(22 c_{12}^2+33 f K_1+16 K_2\right)+8 c_{12}^2 \left(K_2-f K_1\right)
\\ & -8 c_{11}^4+12 f K_1 K_2-6 K_1^2+64 K_2^2)]\,z^4 +  12288 \, c_{11} [256 c_{12} \left(16 c_{11}^2-12 c_{12}^2-27 f K_1+96 K_2\right)
\\ &  +27 \lambda ]\,z^5 +  12288 \, [9 c_{12} \lambda +2048 c_{12}^2 \left(c_{11}^2+2 K_2\right)-192 (64 K_2 \left(c_{11}^2+K_2\right)   \   ,
\\&  -9 K_1^2)-256 c_{12}^4]\,z^6
\end{aligned}
\end{equation*}

\begin{equation*}
\begin{aligned}
\tau_3 \ = & \    -196608 \, \left(128 c_{12} \left(-c_{11}^2+f K_1+4 K_2\right)+\lambda \right)\,z^3
+ 50331648 \,c_{11} \left(-c_{11}^2+2 c_{12}^2+3 f K_1+4 K_2\right)\,z^4
\\ &   -24576\, \left(256 c_{12} \left(32 c_{11}^2-4 c_{12}^2-9 f K_1+32 K_2\right)+9 \lambda \right) \, z^5
+ 50331648 \,c_{11} \left(c_{11}^2-c_{12}^2+6 K_2\right)\,z^6
\end{aligned}
\end{equation*}

\begin{equation*}
\begin{aligned}
\tau_4 \ = & \    -25165824 \, z^4 \left(c_{11}^2 \left(6 z^2-1\right)-c_{12}^2 \left(z^2-2\right)-10 c_{12} c_{11} z+3 f K_1+2 K_2 \left(3 z^2+2\right)\right)    \   ,
\end{aligned}
\end{equation*}

\begin{equation*}
\begin{aligned}
\tau_5 \ = & \ 50331648 \,  z^5 \left(3 c_{11} z-2 c_{12}\right)   \ ,
\end{aligned}
\end{equation*}

\begin{equation*}
\begin{aligned}
\tau_6 \ = & \ -50331648 \, z^6  \  .
\end{aligned}
\end{equation*}

($f=\sqrt{1+z^2}$). For the special values ${K}_1={c}_{12}=0$ and
$K_2=\frac{1}{4} {{c}_{11}}^{\,2}$, the algebraic equation (\ref{AleqA}) becomes

\begin{equation*}
262144 \,z^3 \left(c_{11}-2 T\right){}^6-1024 \, \lambda  \left(9 z^2+8\right) \left(c_{11}-2 T\right){}^3-27 \, \lambda ^2 z \left(z^2+1\right) \ = \ 0 \ .
\end{equation*}

solutions of which coincide with $T_{3,4}$ (\ref{Tcas2}), as it should be.

For the extended Coulomb potential

\[ V(r\,,\theta) \ =\
\frac{a}{r} + \frac{T'(\theta)}{r^2}\ , \]

the corresponding algebraic equation is also a sixth order polynomial equation in $T$ given by

\begin{equation}
\upsilon_0(z) + \upsilon_1(z)\,T(z) + \upsilon_2(z)\,T^2(z)+ \upsilon_3(z)\,T^3(z)+ \upsilon_4(z)\,T^4(z)+ \upsilon_5(z)\,T^5(z)+ \upsilon_6(z)\,T^6(z) \ = \ 0 \  ,
\label{AleqC}
\end{equation}

where $z=\tan \theta$ and

\begin{equation*}
\begin{aligned}
\upsilon_0 \ = & \   128 \left(64 c_{31}^4 \left(K_2-4 f K_1\right)+16 c_{31}^2 \left(f K_2-4 K_1\right){}^2-64 f K_1^3-12 f K_1 K_2^2+K_2^3+48 K_1^2 K_2\right)\,z^2
\\ & -64 c_{31} \left(8 c_{31}^2 \left(64 c_{32} \left(3 f K_1-K_2\right)+\lambda \right)+\left(4 K_1-f K_2\right) \left(64 c_{32} \left(f K_2+5 K_1\right)-9 f \lambda \right)\right)\,z^3
\\ & + (64 [32 c_{32}^2 \left(f^2 K_2^2+10 f K_2 K_1-2 K_1^2\right)-24 c_{31}^2 \left(32 c_{32}^2 \left(2 f K_1-K_2\right)+c_{32} \lambda +24 K_1 \left(f K_2-4 K_1\right)\right)
\\ &  -9 c_{32} f \lambda  \left(f K_2+2 K_1\right)+4 K_2 \left(-6 f K_2 K_1+24 K_1^2+K_2^2\right)]+27 \lambda ^2) \,z^4
\\ &  -128 c_{31} \left(256 c_{32}^3 \left(f K_1-K_2\right)+288 c_{32} K_1 \left(f K_2+2 K_1\right)+12 c_{32}^2 \lambda -27 f \lambda  K_1\right)\,z^5
\\ &  + (27 \left(4096 c_{31}^2 K_1^2+\lambda ^2\right)+128 \left(-4 c_{32}^3 \lambda +64 c_{32}^4 K_2+K_2^3\right))\,z^6      \  ,
\end{aligned}
\end{equation*}

\begin{equation*}
\begin{aligned}
\upsilon_1 \ = & \  -1024\,f\, c_{31} \left(16 c_{31}^2 f \left(K_2-4 f K_1\right)+64 c_{31}^4 f+16 f K_1^2+f K_2^2-8 K_1 K_2\right)\,z^2
\\ &  -32 \,f\, [f^3 K_2 \left(9 \lambda -64 c_{32} K_2\right)-4 K_1 \left(16 c_{32} K_2+9 f^2 \lambda \right)+8 c_{31}^2 \left(64 c_{32} \left(f K_2+11 K_1\right)-15 f \lambda \right)
\\ &   +   1280 \,c_{32}\, f\, K_1^2   +  8192\, c_{32} \,c_{31}^4\, f]\,z^3  + 1024 \,f\,c_{31} [3 c_{32} f \lambda +16 c_{32}^2 f \left(-24 c_{31}^2-16 f K_1+K_2\right)
\\  &  +  96 f K_1 \left(3 c_{31}^2 f+K_1\right)-4 K_2 \left(8 c_{31}^2 f+K_1\right)-4 f K_2^2] \,z^4   +  64\,f\, [-12 c_{32}^2 f \lambda +8 c_{31}^2 \,(9 f \lambda
\\ &  -32 c_{32} (4 f \left(4 c_{32}^2+K_2\right) -7 K_1))  + 256 c_{32}^3 f \left(K_2-f K_1\right)-64 c_{32} K_1 \left(9 f K_1+4 K_2\right)+27 f^2 \lambda  K_1 ] \,z^5
\\ & + 512 \,f\,c_{31} \,\left(f \left(9 c_{32} \lambda -128 c_{32}^4+216 K_1^2\right)-8 K_2 \left(8 c_{32}^2 f+3 K_1\right)-6 f K_2^2\right)\,z^6   \\ &  18432\,f\, c_{32} \,K_1\, \left(16 c_{31}^2-K_2\right)\,z^7  \  ,
\end{aligned}
\end{equation*}

\begin{equation*}
\begin{aligned}
\upsilon_2 \ = & \  256\,f\, \left(16 c_{31}^2 f \left(4 f K_1-K_2\right)-64 c_{31}^4 f-16 f K_1^2-f K_2^2+8 K_1 K_2\right)\,z^2
\\ & + 3072\,f\, c_{31}\, \left(f \lambda -8 c_{32} \left(8 c_{31}^2 f+f K_2+4 K_1\right)\right)\,z^3
\\ &  + 256\,f\, [3 c_{32} f \lambda +32 c_{31}^2 f \left(-44 c_{32}^2+33 f K_1+K_2\right)+16 c_{32}^2 f \left(K_2-16 f K_1\right)+512 c_{31}^4 f+96 f K_1^2
\\ &  -4 f K_2^2-4 K_1 K_2]\,z^4 +   128\,f\, c_{31} \left(64 c_{32} \left(32 c_{31}^2 f-24 c_{32}^2 f-6 f K_2+15 K_1\right)+27 f \lambda \right)\,z^5
\\ & + 128\,f\, \left(9 c_{32} f \lambda +64 c_{32}^2 f \left(16 c_{31}^2-K_2\right)-24 K_2 \left(K_1-8 c_{31}^2 f\right)-128 c_{32}^4 f+216 f K_1^2-6 f K_2^2\right)\,z^6
\\ & 221184\,f\, c_{31}\, c_{32}\, K_1\,z^7\  ,
\end{aligned}
\end{equation*}

\begin{equation*}
\begin{aligned}
\upsilon_3 \ = & \  512\,f\, \left(f \lambda -8 c_{32} \left(8 c_{31}^2 f+f K_2+4 K_1\right)\right)\,z^3 + 8192 c_{31} f^2 \left(8 c_{31}^2-16 c_{32}^2+12 f K_1+K_2\right)\,z^4
\\ &   +  64\,f\, \left(64 c_{32} \left(-2 f \left(4 c_{32}^2+K_2\right)+64 c_{31}^2 f+5 K_1\right)+9 f \lambda \right)\,z^5  -4096 c_{31} f^2 \left(16 c_{31}^2-16 c_{32}^2-3 K_2\right)z^6
\\ & + 36864\,f\, c_{32}\, K_1\,z^7 \  ,
\end{aligned}
\end{equation*}

\begin{equation*}
\begin{aligned}
\upsilon_4 \ = & \ -512\, f^2\, z^4\, \left(16 c_{31}^2 \left(6 z^2-1\right)-16 c_{32}^2 \left(z^2-2\right)-160 c_{32} c_{31} z-24 f K_1-K_2 \left(3 z^2+2\right)\right) \  ,
\end{aligned}
\end{equation*}

\begin{equation*}
\begin{aligned}
\upsilon_5 \ = & -4096\, z^5 \left(3 c_{31} z-2 c_{32}\right) \  ,
\end{aligned}
\end{equation*}

\begin{equation*}
\begin{aligned}
\upsilon_6 \ = & -1024\, z^6 \  .
\end{aligned}
\end{equation*}

For the special values ${K}_1={c}_{32}=0$ and
$K_2=-8\,{{c}_{31}}^{\,2}$, the algebraic equation (\ref{AleqC}) becomes

\[
256 z^3 \left(-36 z^3 \left(2 c_{31}+T\right){}^6-4 \alpha ^3 \left(9 z^2+8\right) \left(2 c_{31}+T\right){}^3+3 \alpha ^6 \left(z^3+z\right)\right) \ = \ 0 \ ,
\]

in agreement with the particular solutions $T_{1,2}$ (\ref{Tcas1}).

\bibliographystyle{aipauth4-1}
\bibliography{sirefst}

\end{document}